\documentclass[twocolumn]{aastex62}

\usepackage{multirow}
\usepackage{array}
\usepackage{enumerate}
\usepackage{amsmath} 

\newcolumntype{P}[1]{>{\centering\arraybackslash}p{#1}}
\received{December 17, 2019}
\revised{February 10, 2020}
\accepted{February 16, 2020}
\published{Astrophysical Journal}
\shorttitle{PV Mixing in a Tangled Magnetic Field}
\shortauthors{C. Chen \& P.H. Diamond}

\begin{document}
\title{Potential Vorticity Mixing in a Tangled Magnetic Field}

\author[0000-0002-0786-7307]{Chang-Chun Chen}
\affil{University of California at San Diego, \\
La Jolla, CA 92093, USA \\
}

\author{Patrick H. Diamond}
\affil{University of California at San Diego, \\
La Jolla, California 92093, USA \\
}

\begin{abstract}
A theory of potential vorticity (PV) mixing in a disordered (tangled) magnetic field is presented. 
The analysis is in the context of $\beta$-plane MHD, with a special focus on the physics of momentum transport in the stably stratified, quasi-2D solar tachocline. 
A physical picture of mean PV evolution by vorticity advection and tilting of magnetic fields is proposed.
In the case of weak-field perturbations, quasi-linear theory predicts that the Reynolds and magnetic stresses balance as turbulence Alfv\'enizes for a larger mean magnetic field. 
Jet formation is explored quantitatively in the mean field--resistivity parameter space. 
However, since even a modest mean magnetic field leads to large magnetic perturbations for large magnetic Reynolds number, the physically relevant case is that of a strong but disordered field. 
We show that numerical calculations indicate that the Reynolds stress is modified well before Alfv\'enization --- i.e. before fluid and magnetic energies balance.
To understand these trends, a double-average model of PV mixing in a stochastic magnetic field is developed. 
Calculations indicate that mean-square fields strongly modify Reynolds stress phase coherence and also induce a magnetic drag on zonal flows. 
The physics of transport reduction by tangled fields is elucidated and linked to the related quench of turbulent resistivity. 
We propose a physical picture of the system as a resisto-elastic medium threaded by a tangled magnetic network. 
Applications of the theory to momentum transport in the tachocline and other systems are discussed in detail.
 

\end{abstract}
\section{Introduction} 
Turbulent momentum transport is a process that plays a central role in the dynamics of astrophysical and geophysical fluids, and in the formation of many astrophysical objects.
Examples of phenomena where momentum transport is at center stage include accretion in both thin and thick disks \citep{balhaw:1998}, the generation of differential rotation in the sun \citep{bretherton1968effect, Spiegel1992, mcintyre2003, Miesch2005} and other stars \citep{Sweet1950, eddington_1988, vainshtein1991}, magnetic dynamos, and atmospheric phenomena in solar system and exoplanets \citep{ingersoll1979zonal, busse1994, maximenko2005observational}. 
Despite the importance of turbulent transport for astrophysics, it is difficult to derive general theories for it. 
Computational models are unable to resolve the vast range of spatial and temporal scales required for a complete description, and also analysis is usually limited. 
However, in certain circumstances, the system can be captured by the development of an asymptotic procedure that represents the essential interactions.

In some cases, the dynamics of the turbulence is effectively two-dimensional (2D) --- usually due to rapid rotation and strong stratification \citep[i.e. small Rossby number and large Richardson number; see, e.g.][] {mcintyre2003}. 
In these cases, it is possible to describe the turbulent dynamics  using classic $\beta$-plane or quasi-geostrophic models \citep{pedlosky1979, bracco1998}, familiar from geophysical fluid dynamics.

The \textit{solar tachocline} is one such quasi-2D astrophysical object \citep{Miesch_2003,Miesch2005, tobias2005}. 
The lower tachocline is a thin, stably stratified layer, thought to sit at the base of the convection zone \citep{basu1997, charbonneau1999, kosovichev1996}, which is of great interest in the context of the solar dynamo \citep{Parker1993, cattaneo1991suppression, Cattaneo1994,wt:2007}, since tachocline shear flows can stretch and so amplify magnetic fields that may be stored there against the action of magnetic buoyancy by the stable stratification \citep{Schou1998}. 
Turbulent transport plays a key role in the tachocline; indeed, it may be responsible for its very existence --- see \citet{Spiegel1992}, \citet{ gough1998inevitability}, and later in the article. 
The nature of the turbulent transport in the tachocline is still uncertain. 
Even such fundamental questions as whether the transport is up or down gradient or significantly anisotropic remain unanswered.

Given the effective 2D structure of the tachocline, it is natural to treat its dynamics using classical shallow water theory and formulate its description in terms of potential vorticity (PV) evolution and transport. 
In the shallow water picture, the PV flux governs the turbulent momentum transport, since the Taylor identity \citep{taylor1915} directly relates the PV flux to the Reynolds force. 
However, the solar tachocline presents additional challenges. 
It is composed of ionized gas and thus must be treated as a magneto-fluid and modeled, for example, by $\beta$-plane or shallow water magnetohydrodynamics (MHD; \citealt{Moffatt1978,  Gilman1997, gilman2000}). 
The tachocline supports a mean azimuthal magnetic field $B_0$. This magnetic field breaks PV conservation in an extremely subtle way \citep{ddt2018}.
Moreover, Rossby waves couple to Alfv\'en waves, so the turbulence has a geostrophic character at some scale, and that of 2D MHD at others. 

Indeed, the plot further thickens. 
The solar tachocline is strongly forced by convective overshoot from the convection zone. 
Thus, the magnetic Reynolds number ($Rm \equiv VL/\eta$, where $\eta$ is resistivity) is large. 
From the Zel'dovich relation for 2D MHD \citep{fyfe_montgomery_1976, Gruzinov1996, Diamond2005}, we can expect the root-mean-square (rms) magnetic field $\langle \widetilde{B} \rangle^{1/2}$ to vastly exceed the mean field $B_0$ in the tachocline, where angle brackets $\langle\;\rangle \equiv \frac{1}{L} \int dx \frac{1}{T} \int dt$ represent the ensemble average over long space scales and timescales, and $\widetilde{\;}$ denotes perturbations vary away from the mean. 
Thus, though the tachocline is surely magnetized, its field is neither smooth nor uniform. 
This points to the topic of \textit{PV transport in a tangled field} --- the subject of this paper --- being crucial for understanding momentum transport in the tachocline.
 
Previous studies of flow dynamics for $\beta$-plane MHD have focused on PV transport and jet (zonal flow) formation \citep{diamond2007beta, hughes_rosner_weiss_2007, Diamond2005,leprovost2007effect}. 
Computational studies have noted that even weak mean magnetic fields can inhibit negative viscosity phenomena such as jet formation \citep{Miesch_2001,Miesch_2003, Tobias2007, G_rcan_2015}.
Results indicate that for fixed forcing and dissipation, jets form for $B_0^2/\eta < (B_0^2/\eta)_{crit} $, but are inhibited for $B_0^2/\eta > (B_0^2/\eta)_{crit}$. 
These findings are interpreted in terms of the classical idea that the mean field, $B_0$, tends to `Alfv\'enize' the turbulence , i.e. converts Rossby wave turbulence to Alfv\'en waves turbulence. 
For Alfv\'enic turbulence, fluid and magnetic stresses tend to compete, thus restricting PV mixing and inhibiting zonal flow formation \citep{Diamond_2005}. 
When the freezing-in law \citep{poincare1893} is not violated, the strong field--fluid coupling prevents PV mixing and (loosely put) the 2D inverse energy cascade \citep{Kraichnan1965,Iroshnikov1964,Biskamp1989}. 
When irreversible resistive diffusion is sufficiently large to break freezing-in, PV mixing occurs. 

As noted earlier, these fundamental issues are of great relevance to the tachocline, since momentum transport is vital to its formation. 
Specifically, the tachocline may be thought to form by `burrowing' driven by large meridional cells. These, in turn, are driven by baroclinic torque (i.e. $\nabla p \times \nabla \rho$; \citealt{Mestel1999}). 
In one leading model --- that of \citet{Spiegel1992} --- burrowing is opposed by turbulent viscous diffusion of momentum in latitude.
In another model --- proposed by \citet{Gough1998} --- burrowing is opposed by PV mixing and by a hypothetical fossil magnetic field in the solar radiation zone. 

The \citet{Spiegel1992} model ignores the true nature of 2D tachocline dynamics. 
\citet{Gough1998} ignore the effect of magnetic fields in turbulent momentum transport and the implication of Alfv\'en's theorem. 
Neither tackles the strong stochasticity of the ambient tachocline field.
Recent progress on this subject has exploited theoretical approaches based on quasi-linear (QL) theory or wave turbulence theory \citep{cp2018}.
These are unable to take into account for the stochasticity of the ambient field; i.e. the fact that $|\widetilde{B}^2|/ B_0^2 \gg 1$ in the tachocline, where fields  are strongly tangled. 

One indication of the deficiency in the conventional wisdom is the observation from theory and computation that values of $B_0^2$ well \textit{below} that for Alfv\'enization are sufficient to ensure the reduction in Reynold stress and thus PV mixing \citep{Field_2002, Mininni2005, Tobias2007, SILVERS2005, silver2006, Keating_Diamond2007, Keating_2008, keating_diamond_2008, Eyink2011, Kondi__2016, mak2017}. 
This suggests that tangled magnetic fields act to reduce the phase correlation between $\widetilde{u}_x$ and $\widetilde{u}_y$ in the turbulent Reynolds stress $\langle \widetilde{u}_y \widetilde{u}_x \rangle$. 
Note that, as we will show here,  this effect is one of dephasing, not suppression, and not due to a reduction of turbulence intensity. 
It resembles the well-known effect of quenching of turbulent resistivity in 2D MHD, which occurs for weak $B_0^2$ but large $\langle \widetilde{B}^2 \rangle$ (i.e. large $Rm$), at fixed drive and dissipation \citep{cattaneo1991suppression,Cattaneo1994}. 
Thus, it appears that Alfv\'enization --- in the usual sense of  the $\rho_0 \langle \widetilde{v}^2 \rangle = \langle \widetilde{B}^2 \rangle/\mu_0$ balance intrinsic to linear Alfv\'en waves --- and the associated stress cancelation are \textit{not} responsible for the inhibition of PV mixing in $\beta$-plane MHD at high magnetic Reynolds number. 
This observation reinforces the need to revisit the problem with a fresh approach. 

In this paper, we present a theory of PV mixing in $\beta$-plane MHD. 
A mean field theory is developed for the weak perturbation regime, and a novel model is derived for the case of a strong tangled field ($\langle \widetilde{B}^2\rangle > B_0^2$).
The latter is rendered tractable by considering the fluid dynamics to occur in a prescribed static, stochastic field. 
For $\langle \widetilde{B}^2\rangle < B_0^2$, the quasi-linear calculation reveals that PV mixing evolves by both advection and by inhomogeneous tilting of field lines correlated with fluctuations. 
The presence of $B_0$ converts Rossby waves to Rossby-Alfv\'en waves so the system exhibits a stronger Alfv\'enic character for larger $B_0$. 
When turbulence Alfv\'enizes, PV mixing is quenched by the balance of fluid and magnetic stresses.
However, the issue is more subtle, since numerical calculations reported here indicate that \textit{magnetic fields affect the Reynolds stress well before the point of Alfv\'enization}.
This suggests that magnetic fluctuations affect the phase \textit{correlation} of velocity fluctuation in the stress, in addition to producing the competing magnetic stress. 
By the Zel'dovich theorem, however, we expect that $|\widetilde{B}^2| \gg B_0^2$, so QL theory formally fails. 
To address the $|\widetilde{B}^2| \gg B_0^2$ limit, we go beyond QL theory and consider an effective medium theory, 
which allows calculation of PV mixing in a resisto-elastic fluid, where the elasticity is due to $\langle \widetilde{B}^2\rangle$.
The resisto-elasticity of the system acts to reduce the phase correlation in the Reynolds stress. 
Physically, fluid energy is coupled to damped waves, propagating through a disordered magnetic network. 
The dissipative nature of the wave-field coupling induces a drag on the mesoscale flows.
We show that PV mixing is quenched at large $Rm$, for even a weak $B_0$. 
The implications for momentum transport in the solar tachocline and related problems are discussed.  

The remainder of this paper is organized as follows. 
Section \ref{sec: Models} presents and elucidates models and the QL theory of $\beta$-plane MHD. 
Section \ref{sec: TangledModel} details the effective medium theory of PV mixing in a tangled magnetic field. 
The phase correlation in the Reynolds stress and the onset of magnetic drag are calculated. 
A physical model of the effective resisto-elastic medium is discussed. 
Section \ref{sec: discussion and conclusion} presents the conclusions and discusses the application of the theory, along with future work.

\section{Models} \label{sec: Models}
In this Section, we present the $\beta$-plane MHD model and discuss its relevance to the solar tachocline. 
The physics of PV transport in $\beta$-plane MHD is described. 
Both mixing by fluid advection and magnetic tilting are accounted for. 
\subsection{Zonal Flow and PV Mixing in $\beta$-Plane MHD Model}\label{sec: MHD beta plane}
\begin{figure}[h!]\centering
  	\includegraphics[scale=0.3]{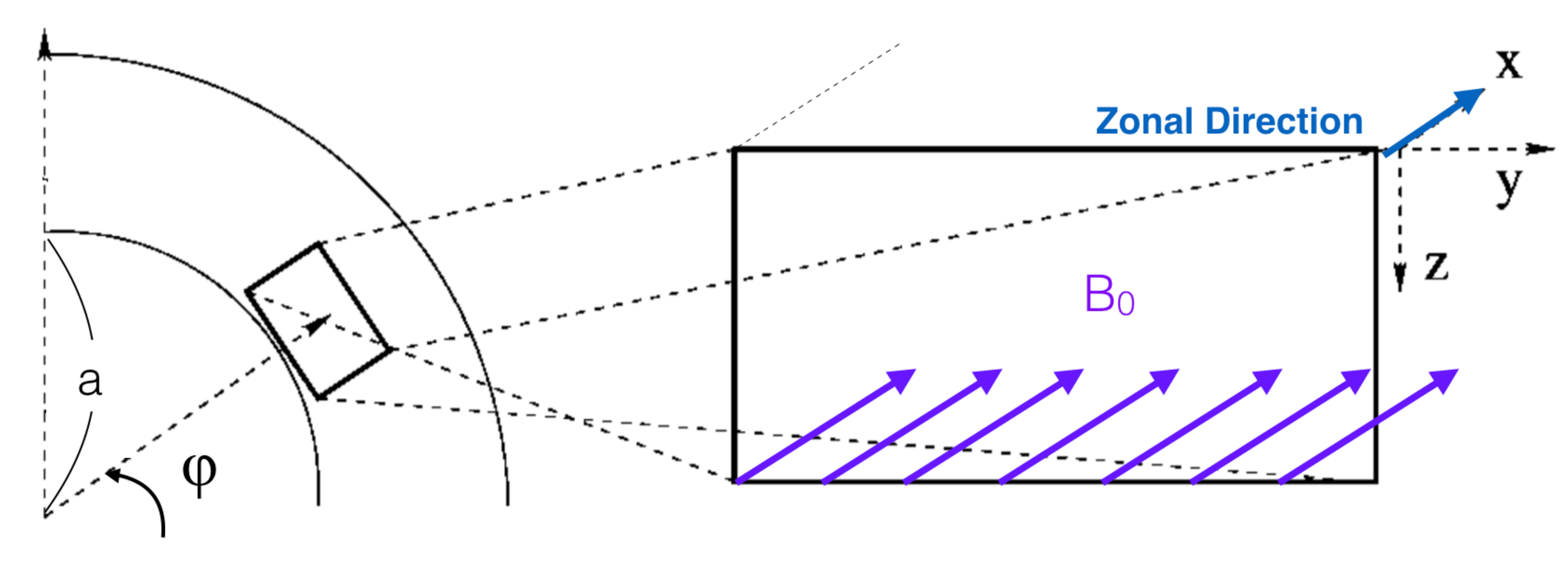}
  	\caption{Geometry and computational domain for the local Cartesian model. 
	$x$- and $y$-axis are local longitudinal and latitudinal directions, respectively.
	The $z$- axis represents the depth of the $\beta$-plane.
	The mean magnetic field $B_0$ is zonal direction ($x$-axis). 
	}
  	\label{fig: beta-plane model}
\end{figure}

The solar tachocline is a thin layer inside the Sun, located at a radius of at most 0.7 $R_{\odot}$, with a thickness of $\lesssim 0.04$ $R_{\odot}$ \citep{cdthom2007}.
Dynamics on this thin shell can be modeled using the $\beta$-plane, following a model proposed by \citet{rossby1939relation}, for the thin atmosphere. 
In this model, $\beta$ is defined as the  \textbf{Rossby Parameter}, given by  
$
	\beta = \frac{df}{dy}|_{\phi_0} = 2\Omega cos(\phi_0) /a.
$
Here, the $f \equiv 2\Omega \sin \phi_0 + \beta y$ is the angular frequency at latitude $\phi_0$ on the $\beta$-plane, $y$ is the meridional distance from $\phi_0$, and $\Omega = | \bold{\Omega}|$ is the angular rotation rate of the planet. 
The angular frequency $f$ is also known as the \textbf{Coriolis parameter}. 
Notice that $\phi_0$ increases from the equator (see Figure \ref{fig: beta-plane model}).
The simplified $\beta$-plane MHD model extends the hydrodynamic model to include the effects of MHD and comprises two basic scalar equations:
\begin{align}
	 	\left(\frac{\partial}{\partial t} + \bold{u} \cdot \nabla  \right)\zeta -\beta\frac{\partial\psi}{\partial x} 
				&= -\frac{(\bold{B}\cdot\nabla)(\nabla^2 A)}{\mu_0 \rho} + \nu \, \nabla^2 \zeta,
		\label{eq:scalar1}
\\
	  \frac{\partial}{\partial t} A 
	& = (\bold{B} \cdot \nabla) \psi + \eta\nabla^2 A.
	\label{eq:scalar2}
\end{align}
These two scalar equations are from the Navier-Stokes equation and the induction equation, respectively.
Here, $\eta$, $\mu_0$, and $\rho$ are the magnetic diffusivity, the permeability, and the density, respectively. 
The scalar $\psi$ is the $z$-component of the stream function $\bold{\Psi} = (0, 0, \psi)$ for 2D incompressible flow, so that $\bold{u} = (\frac{\partial}{\partial y} \psi, -\frac{\partial}{\partial x}  \psi, 0)$, and $A$ is the scalar potential for the magnetic field $\bold{A}= (0, 0, A)$. We also define the vorticity $\zeta \equiv -\nabla^2 \psi$, similar to the relationship between the current and the potential $ J \equiv - \mu_0\nabla^2 A$.
Eq. \ref{eq:scalar1} and \ref{eq:scalar2} show that the vorticity and the potential field $A$ are conserved in $\beta$-plane, up to the Lorentz force, resistivity, and viscosity. 

\begin{figure}[h!]\centering
  	\includegraphics[scale=0.44]{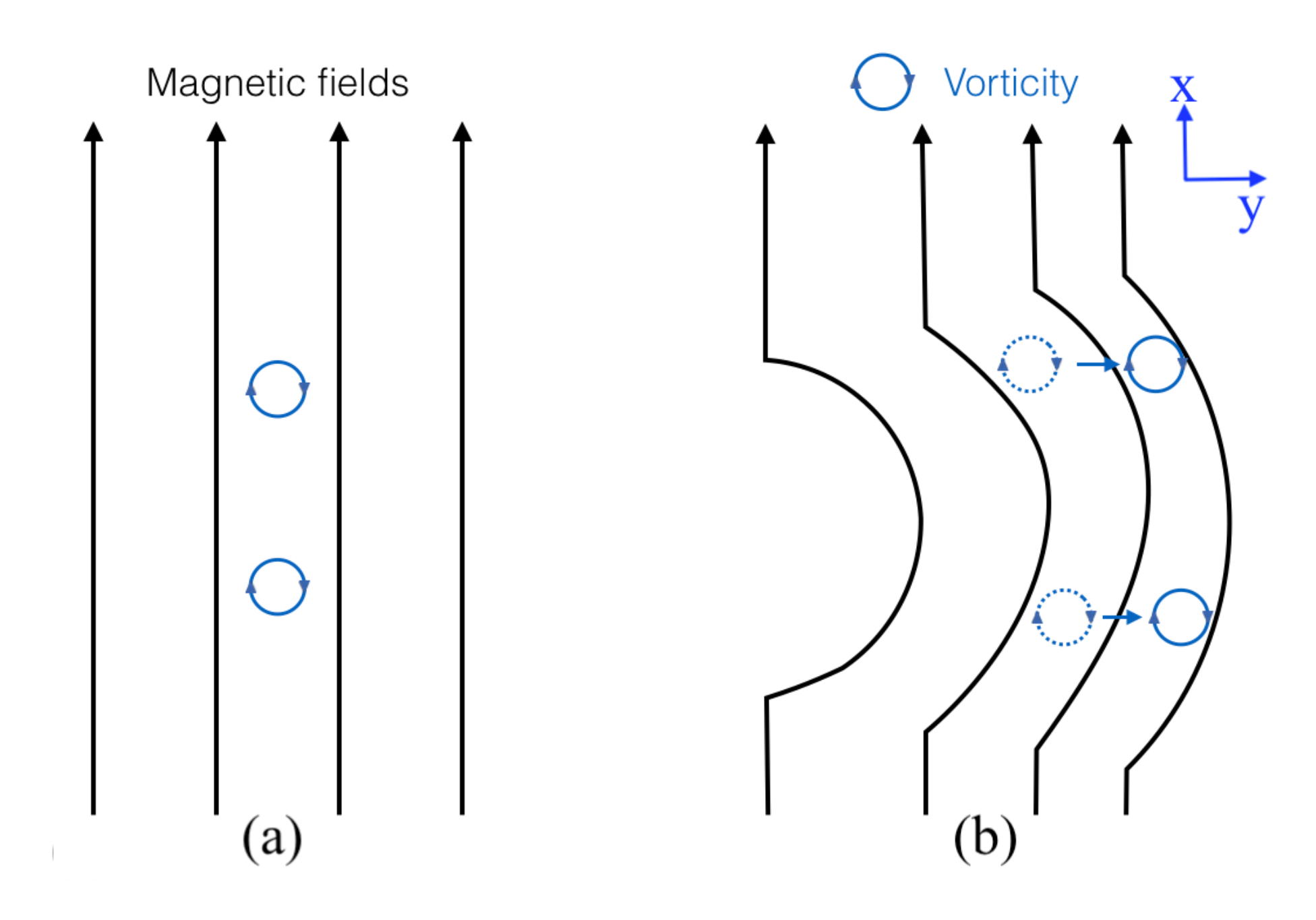}
  	\caption{Evolution of PV threaded by magnetic field lines in a frame moving with the flow.
	Aside from the advection of flow, the distribution of PV charge density also changed under the influence of inhomogeneous magnetic fields.  (a) PV uniformly distributed in the moving frame. (b) PV distribution is changed by the tilted magnetic field lines. Dashed circles are undisturbed vortices. Solid circles are new locations of PV charge density. 
	}
  	\label{fig:tilted_field_vorti}
\end{figure}

The 2D hydrodynamic inviscid shallow water equation illustrates physics of the solar tachocline. 
The \textbf{PV Freezing-in Law} describes how the PV is frozen into the fluid.
In $\beta$-plane model, the generalized PV that frozen into fluid is the potential vorticity $PV \equiv \zeta + f$, where $\zeta$ is the vorticity as defined, and $f$ is the Coriolis parameter.
This freezing-in of the PV is broken by body forces, such as the Lorentz force, and by the viscosity.
To illustrate how the PV freezing-in law is broken, we first split the parameters into two parts, representing two-scale dependences.
The shorter length is the turbulence wavelength and the longer length is the scale over which we perform the spatial average.
Applying this mean field theory to Eq. \ref{eq:scalar1} and \ref{eq:scalar2} leads to: 
\begin{equation}
\frac{D}{D t}  \langle \zeta \rangle  
			=  \frac{\partial}{\partial y}\frac{\langle \widetilde{J}_{z} \widetilde{B}_{y} \rangle }{ \rho} 
			+ \nu \nabla^2 \langle \zeta \rangle
			\neq 0.
		\label{eq: freezing-in beak by Lorentz}
\end{equation}
In this form, we can interpret PV density as a `charge density element' ($ \zeta\equiv \rho_{PV}$), 
floating in the fluid threaded by stretched magnetic fields (see Figure \ref{fig:tilted_field_vorti}).
Using charge continuity 
\[
\frac{\partial}{\partial t} \rho_{PV} + \nabla \cdot \bold{J}_{PV} = 0, 
\]
and writing the current as 
\[
\bold{J}_{PV} = \bold{J}_{PV, \parallel} + \bold{J}_{PV, \perp}, 
\]
we have 
\[
\frac{\partial}{\partial t} \rho_{PV} = -\nabla_{\perp} \cdot \bold{J}_{PV, \perp} - \nabla_{\parallel} \cdot \bold{J}_{PV, \parallel}.
\]
Here, perpendicular and parallel current are 
\[
\bold{J}_{PV, \perp} = \bold{v}_{\perp} \rho_{PV}\quad {\rm and} \quad \bold{J}_{PV, \parallel} = \frac{\bold{B}}{|B|} J_{PV, \parallel},
\]
respectively. 
The direction parallel ($_{\parallel}$) and perpendicular ($_{\perp}$) to the mean magnetic field are along the $x$- and $y$-axis, respectively. 
We stress here that $\nabla_{\perp}$ is non-linear ($\nabla_{\perp} = \frac{\partial}{\partial y}  +\frac{\widetilde{B}}{B_0} \frac{\partial}{\partial y} $).
Thus: 
\begin{equation}
	\frac{\partial}{\partial t} \langle \rho_{PV} \rangle 
	= -\frac{\partial}{\partial {y}} \langle \widetilde{u}_{y} \widetilde{\rho}_{PV} \rangle 
	+ \frac{\partial}{\partial y}\frac{ \langle \widetilde{B}_{y}  \widetilde{J}  \rangle}{ \rho}
	+\nu \nabla^2 \langle \rho_{PV} \rangle .
	\label{eqn: charge density}
\end{equation}
The first term in Eq. \ref{eqn: charge density} is the contribution to the change in charge density from the divergence of the latitudinal flux of vorticity, while the second term is the contribution because of the inhomogeneous tilting of the magnetic field lines.

Figure \ref{fig:tilted_field_vorti} shows a cartoon of how the PV charge density is related to `plucking' magnetic lines.
In $\beta$-plane MHD, zonal flows are produced by inhomogeneous PV mixing (i.e. an inhomogeneous flux of PV `charge density') and by the inhomogeneous tilting of magnetic field lines (weighted by current density).
In simple words, there are two ways to redistribute the charge density (in this case, the absolute vorticity) --- one is through advection, and the other is by bending the magnetic field lines, along which current flows. 
These two processes together determine the net change in local PV charge density.

A second tool that can be brought to bear on understanding of the physics of PV mixing is the \textbf{Taylor identity} \citep[ $\langle \widetilde{u}_y\widetilde{\zeta}\rangle 
				=  - \frac{\partial}{\partial y} \langle \widetilde{u}_y\widetilde{u}_x\rangle$; see][]{taylor1915}.
This can be extended to the 2D MHD case by deriving the \textbf{extended Taylor identity}, useful in the context involves the Maxwell stresses. 
We begin with two scalar fields decomposed as: 
\begin{align}
	\zeta &= \langle \zeta \rangle + \widetilde{\zeta},
	\nonumber
	\\
	A &= \langle A \rangle + \widetilde{A}.
\end{align}
Again, scalar fields $\widetilde{\zeta}$ and $\widetilde{A}$  represent the perturbations of vorticity and the potential field, respectively, due to waves and turbulence.  
For the hydrodynamic case, use of the Taylor identity, which relates the vorticity flux to the Reynolds force, leads to the derivation of  the zonal flow evolution equation:
\[
	\frac{\partial}{\partial t}  \langle u_x \rangle 
		 		= \langle \widetilde{u}_y\widetilde{\zeta}\rangle  
				= - \frac{\partial}{\partial y} \langle \widetilde{u}_y\widetilde{u}_x\rangle.
				\label{eqn:PVflux}
\]
This equation shows that the cross-flow flux of potential velocity underpins the Reynolds stress and that the gradient of the Reynolds stress (a shear force) then drives the large-scale zonal flow. 
The link between inhomogeneous, cross-flow PV transport (i.e. PV mixing), and mean flow generation is established. 

We introduce the extended Taylor identity --- an analogous form for the magnetic field perturbations in MHD:
\[
	  \frac{\langle \widetilde{B}_y \nabla^2 \widetilde{A}\rangle}{\mu_0}	
				=   -  \langle \widetilde{B}_y \widetilde{J} \rangle
				=   \frac{\partial}{\partial y} \frac{\langle \widetilde{B}_y\widetilde{B}_x\rangle}{\mu_0},
	\label{eq:ext-taylar-identity}
\]
and therefore
\begin{equation}
\frac{\partial}{\partial t}  \langle u_x\rangle  = - \frac{\partial}{\partial y}	
						\bigg\{ \langle \widetilde{u}_{x}\widetilde{u}_{y} \rangle
	 	   	-  \frac{\langle \widetilde{B}_{x}\widetilde{B}_{y} \rangle }{\mu_0 \rho}\bigg\}
			+ \nu \nabla^2 \langle u_x \rangle.
			\label{eq: reynolds and maxwell stresses competition}	
\end{equation} 
This equation states that the mean PV transport is determined by the difference between the Reynolds and Maxwell stresses. 
In a perfectly Alfv\'enized state, the total momentum flux vanishes, owing to the cancellation of the Reynolds and Maxwell stresses ($\rho \langle \widetilde{u}^2 \rangle = \langle \widetilde{B}^2 \rangle/\mu_0$).
\subsection{Validity of QL Theory} \label{subsec: Validity of QLT}

We start by analytically deriving the mean PV flux using QL theory.
This calculation employs the linear responses of vorticity and magnetic potential fields to estimate the evolution and the relaxation of the flow. 
Similar calculation can be found in the QL closure done by \citet{pouquet_1978} and \citet{ McComb1990}.
Before presenting the QL calculation, we first discuss its validity.  
The key to the latter is the dimensionless parameter --- the Kubo numbers ($Ku$; \citealt{kubo1963}) --- that quantifies the effective memory of the flow and the field. 

\begin{figure}[h!]\centering
  	\includegraphics[scale=0.46]{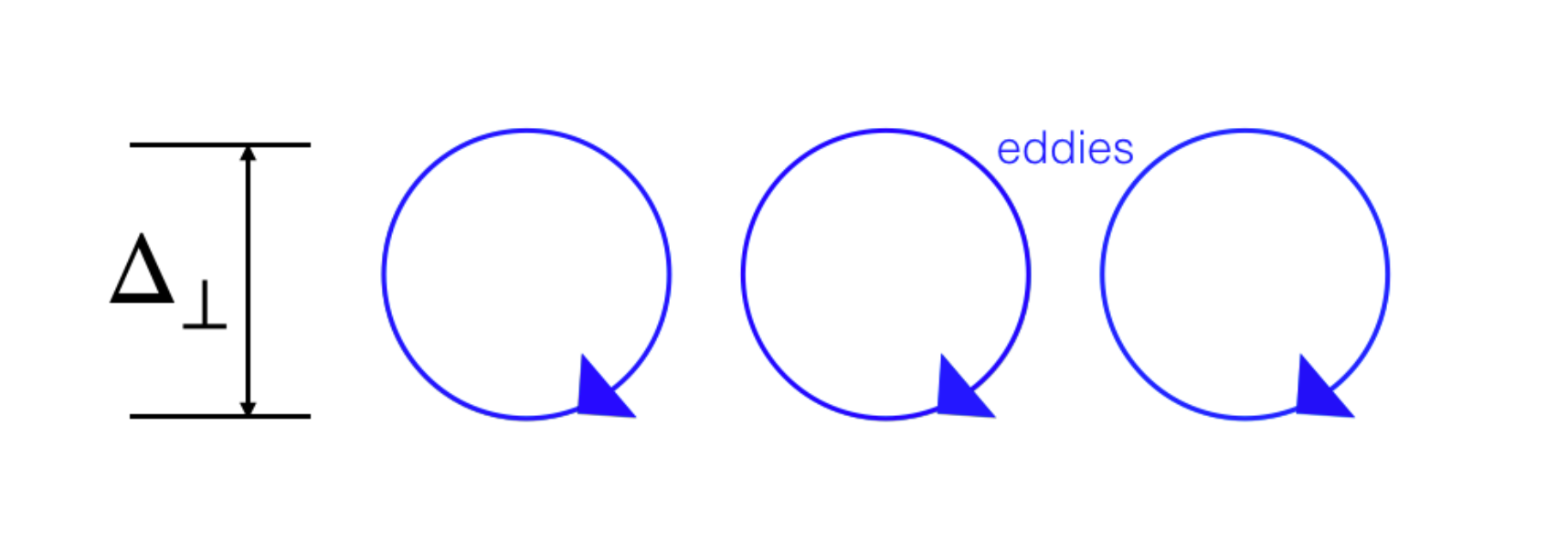}
  	\caption{Eddy size $\Delta_{\perp}$. In this figure, the shear flow is in the left-right direction.  The eddy size is measured perpendicular to the flow. 
	}
  	\label{fig:eddies}
\end{figure}

The \textbf{fluid Kubo number} is defined as 
\begin{equation}
	Ku_{fluid}  \equiv \frac{ \delta_l }{\Delta_{\perp}} 
	\sim \frac{\widetilde{u} \tau_{ac} }{\Delta_{\perp}}
	\sim \frac{\tau_{ac}}{\tau_{eddy}},
\end{equation}
where $\delta_l $ is the characteristic scattering length, $\tau_{ac}$ is the velocity autocorrelation time, and $\tau_{eddy}$ is the eddy turn-over time. 
The eddy turn-over time is $
	\tau_{eddy} = \Delta_{\perp}/\widetilde{u} ,
$
where $\Delta_{\perp}$ is the eddy size (see Figure \ref{fig:eddies}). 
In practice, the validity of QL theory requires small fluid Kubo number $Ku \ll 1$.
To understand this, we compare autocorrelation rate ($1/\tau_{ac}
	\equiv \Delta \left(-\beta k_x/k^2\right) 
	= \left|-\beta/k^2  + 2\beta k_x^2/k^4 \right| \Delta k_x + \left|2\beta k_x k_y / k^4 \right|\Delta k_y$) with decorrelation rate ($1/\tau_{eddy} = k \widetilde{u}$) on $\beta$-plane.  
This gives $\tau_{ac} < \tau_{eddy} $ (or equivalently $\l_{ac} < \Delta_{\perp}$), leading to $Ku_{fluid} < 1$.
As a particle traverses an eddy length, it experiences several random kicks by the flow perturbations, as in a diffusion process. 
In this limit, trajectories of particles don't deviate significantly from unperturbed trajectories. 
Note that in the case of wave turbulence, the autocorrelation time ($\tau_{ac}$) is sensitive to dispersion.
The autocorrelation time can be expressed as: 
\begin{equation}
	\frac{1}{\tau_{ac}} = \Delta \omega = \frac{d\omega}{d\bold{k}} \cdot \Delta \bold{k} .
\end{equation} 
However, when the turbulence is strong, we have $\delta_l \gg \Delta_{\perp}$. 
Here, particles deviate strongly from the original trajectories in an autocorrelation time, indicating a failure of QL theory (i.e. $Ku_{fluid}  >1$). 
However, it is clear that $\beta$-plane MHD is not a purely fluid system; hence the validity of QL theory depends not only on the fluid Kubo number but also on the \textbf{magnetic Kubo number}.
This can be written as:
\begin{eqnarray}
	Ku_{mag}  &\equiv \dfrac{\delta_{l}}{\Delta_{\perp}} 
		\\
	\delta_l &\sim \dfrac{l_{ac}|\widetilde{  \bold{B}} | }{B_0},
\end{eqnarray}
where $\delta_l$ is the deviation of a field line, $l_{ac}$ is the magnetic autocorrelation length, and $|\widetilde{B}|$ is the magnetic field intensity of the wave turbulence.  
If a particle travels a coherence length $\Delta_{\perp}$ and experiences several random kicks in weak magnetic perturbations, it undergoes a process of magnetic diffusion, which can be treated using QL theory \citep{Rosenbluth1977}.
In contrast, when magnetic perturbations are strong, particle trajectories are sharply deflected by strong $\widetilde{B}$-induced scattering within an autocorrelation length. 

Our main interest in this paper is the case in the solar tachocline, where zonal flows and eddies coexist, $Rm$ is large, and the magnetic field lines are strongly stretched and distorted by the turbulence. 
Hence, the fluid Kubo number is modest (i.e. $Ku_{fluid} \lesssim 1$), and the magnetic Kubo number is small  $Ku_{mag} \ll 1$ (see Table \ref{tab:1}).
This is done by taking small-scale fields as spatially uncorrelated ($l_{ac} \rightarrow 0$).  
Details are discussed in Section \ref{sec: TangledModel}.
\setcounter{table}{0}
\begin{table}[h!]
\renewcommand{\thetable}{\arabic{table}}
\centering
\begin{tabular}{P{0.13\textwidth}>{\centering} P{0.15\textwidth} p{0.15\textwidth} }
\tablewidth{0pt}
\hline
\hline
 &Fluid & Magnetic  \\
\hline
\decimals
Operator & $\textbf{u}$ $\cdot  \nabla$ &  $\widetilde{B}/B_0$ $\textbf{u}_A \cdot  \nabla$  \\
Ratio  & $\delta_l /\Delta_{\perp}$$\lesssim 1$ &  $\widetilde{B}/B_0\gg 1$ \\
\multirow{ 2}{*}{For QL Theory } & $\tau_{ac}$   $\rightarrow 0$ & $l_{ac}$   $\rightarrow$ 0   \\
Validity & (delta-correlated flows) & (uncorrelated tangled fields)\\
Kubo number in the model &  $Ku_{fluid} $  $\lesssim 1$  & $Ku_{mag}  $ $\ll 1$ \\
\hline
\end{tabular}
\par
\bigskip
\caption{
Summary of the properties of fluid and magnetic Kubo numbers. 
All models in this paper are set up to make Kubo numbers small to ensure the QLT is valid. 
This is fulfilled by assuming flows and fields are delta correlated in time and space, respectively.} 
\label{tab:1}
\end{table}
\subsection{Mean Field Theory for $\beta$-Plane MHD} \label{sec:APS_Results}
We first consider the simple case where the large-scale magnetic field $B_0$ is stronger than the small-scale magnetic fields (i.e. $|\widetilde{B}^2|/B_0^2 \ll 1 $). 
Here, the fluid turbulence is weak (restricted by $B_0$), and the tilt of the magnetic field lines are small, corresponding to a small magnetic Kubo number.
To construct the QL equations, we linearize Eq. \ref{eq:scalar1} and \ref{eq:scalar2}:
\begin{eqnarray}
	&\dfrac{\partial}{\partial t}  \widetilde{ \zeta} 
	+ \widetilde{ u}_y \dfrac{\partial \langle \zeta \rangle}{\partial y} 
	+\beta \widetilde{u}_{y}  
	= 
	-\dfrac{B_0}{\mu_0 \rho} \dfrac{\partial (\nabla^2 \widetilde{A})}{\partial x} + \nu \nabla^2 \widetilde{\zeta}
\\
	&\dfrac{\partial}{\partial t}  \widetilde{A} = B_0 \widetilde{ u}_y + \eta\nabla^2 \widetilde{A},
\end{eqnarray}
and obtain the linear responses of vorticity and magnetic potential at wavenumber $k_x$ in the zonal direction to be:
\begin{eqnarray}
		&\widetilde{\zeta}_k  = - \bigg(\dfrac{i}{ \omega+i\nu k^2 +\big(\frac{-B_0^2}{\mu_0 \rho}\big)\frac{ k_x^2}{\omega+i\eta k^2 } }\bigg)  \big(\widetilde{u}_y \dfrac{\partial}{\partial y} \langle \zeta \rangle + \beta\widetilde{u}_y \big),
		\nonumber
\label{linear_resp_vorticity}
\\
		&\widetilde{A}_k = \dfrac{\widetilde{\zeta}_k}{k^2} \big( \dfrac{B_0 k_x}{-\omega-i\eta k^2 } \big),
		\nonumber
\end{eqnarray}
where $k \equiv k_x^2 + k_y^2$.
From these, the dispersion relation for the ideal Rossby-Alfv\'en wave follows:
\begin{equation}
	 \bigg(  \omega - \omega_R   + i\nu k^2  \bigg) 
	 \bigg( \omega  
	 +i \eta k^2 \bigg) 
	 = \omega_A^2.
\label{eq: dispersion}
\end{equation}
Here $\omega_A$ is \textbf{Alfv\'en frequency} ($\omega_A \equiv B_0 k_x/\sqrt{\mu_0 \rho}$), and $\omega_R$ is \textbf{Rossby frequency} ($\omega_R \equiv -\beta k_x/k^2$).
We also derive the QL evolution equation for mean vorticity:
\begin{equation} 
		 	\frac{\partial}{\partial t}\langle \zeta \rangle  
			= -\frac{\partial}{\partial y} \bigg(
			\langle \widetilde{u}_y \widetilde{\zeta} \rangle 
	 			+ \frac{\langle \widetilde{B}_{y}\nabla^2\widetilde{A} \rangle}{ \mu_0 \rho}
				 \bigg) 
				+ \nu \nabla^2 \langle \zeta \rangle.
	\label{eq:evol_vorticity}
\end{equation} 
Using the Taylor identity, the averaged PV flux ($\langle \Gamma \rangle \equiv \langle \widetilde{u}_y \widetilde{\zeta} \rangle + \langle \widetilde{B}_{y}\nabla^2\widetilde{A} \rangle/ \mu_0 \rho$) can be expressed with two coefficients, the fluid and magnetic diffusivities ($D_{fluid}$ and $D_{mag}$):
\begin{equation}
	\dfrac{\partial}{\partial t} \langle \zeta \rangle
	= -\dfrac{\partial}{\partial y} \langle \Gamma \rangle 
	\equiv-\dfrac{\partial}{\partial y} \bigg( - (D_{fluid} - D_{mag})   \frac{\partial}{\partial y} \langle PV \rangle
	\bigg)
	,
	\label{eq: PV flux and Diffusivities}
\end{equation}
Note two aspects of Eq. \ref{eq: PV flux and Diffusivities}.
One is that the anisotropy and inhomogeniety of vorticity flux (i.e. $\frac{\partial}{\partial y}  \langle \zeta \rangle$) leads to the formation of zonal flow. 
For a not-fully-Alfv\'enized case ($D_{mag} < D_{fluid}$), 
zero PV transport occurs when $\frac{\partial}{\partial y}  \langle \zeta \rangle = -\beta$.
This states that $\beta$ provides the symmetry breaking necessary to define zonal flow orientation.
The second aspect is the well-known competition between Reynolds and Maxwell stresses that determines the total zonal flow production. 
These two diffusivities are related to the  Reynolds and Maxwell stress by:
\begin{eqnarray}
   	D_{fluid} \frac{\partial}{\partial y} PV &=   \dfrac{\partial}{\partial y}\langle\widetilde{u}_x\widetilde{u}_y\rangle
	\\
	D_{mag} \frac{\partial}{\partial y} PV &= \dfrac{\partial}{\partial y}\dfrac{\langle\widetilde{B}_x\widetilde{B}_y\rangle}{\mu_0\rho}.
\end{eqnarray}
To calculate the turbulent diffusivities, we express terms $ \widetilde{u}_y \widetilde{\zeta}$ and $\widetilde{B}_{y, \, k}\nabla^2\widetilde{A}_k$ in Eq. \ref{eq:evol_vorticity}  as summations over components in the k space, i.e. $ \widetilde{u}_y \widetilde{\zeta} = \sum_k \widetilde{u}_{y,\, k}^* \widetilde{\zeta}_k$. 
Thus, from Eq. \ref{linear_resp_vorticity} : 
\begin{align}
	& \widetilde{u}_{y, \, k}^*\widetilde{\zeta}_k  
				= \bigg(\dfrac{-i}{ \omega+i\nu k^2 + \frac{-B_0^2}{\mu_0 \rho}\frac{ k_x^2}{\omega +i\eta k^2} }\bigg)  	
				|\widetilde{u}_y|^2 \dfrac{\partial}{\partial y}PV,
\\
			& \widetilde{B}_{y, \, k}^* \nabla^2  \widetilde{A}_k
				= \big(\dfrac{-B_0^2 k_x^2}{\omega^2 +\eta^2 k^4}\big)   \widetilde{u}_{y, \, k}^*\widetilde{\zeta}_k .
	\label{vorti_flux}
\end{align}
Equation \ref{vorti_flux} links the magnetic and fluid diffusivities such that 
\[
 D_{mag} \frac{\partial}{\partial y}  PV  = \frac{1}{\mu_0\rho}\big(\frac{B_0^2 k_x^2}{\omega^2 +\eta^2 k^4}\big) D_{fluid} 
				\frac{\partial}{\partial y}   PV
,
\]
leading to 
\begin{equation}
D_{mag} =   \frac{1}{\mu_0\rho}\left(\frac{B_0^2 k_x^2}{\omega^2 +\eta^2 k^4}\right) D_{fluid}
.
\end{equation}
Hence,
\begin{eqnarray}
	&D_{fluid} =  \sum\limits_{k} C_{k,\, fluid} |\widetilde{u}_{y,\, k}|^2 
		\nonumber \\
	&D_{mag} =  \dfrac{1}{\mu_0\rho} \sum\limits_{k}  C_{k,\, mag} |\widetilde{u}_{y, \, k}|^2,
	\nonumber
\end{eqnarray}
where the phase coherence coefficients $C_k$ are given by
\begin{align}
	&C_{k,\, fluid} = \dfrac{\nu k^2 
		+\frac{ \omega_A^2 \eta k^2} {\omega^2 + \eta^2 k^4} }
		{\omega^2\big(1 - \frac{\omega_A^2}{\omega^2+\eta^2k^4} \big)^2 +\big( \nu k^2 + \omega_A^2 \frac{\eta k^2  }{\omega^2+\eta^2k^4}\big)^2 },
		\label{eq: fluid_PV_diffusivity} 
\\
	&C_{k,\, mag} = \dfrac{\omega_A^2  \big( \frac{\nu k^2}{\omega^2+\eta^2k^4} + \frac{\omega_A^2\eta k^2}{(\omega^2+\eta^2k^4)^2}  \big)^2}
			{\omega^2\big(1 - \frac{\omega_A^2}{\omega^2+\eta^2k^4} \big)^2 +\big( \nu k^2 + \omega_A^2 \frac{\eta k^2  }{\omega^2+\eta^2k^4}\big)^2 }.
		\label{eq: mag_PV_diffusivity}
\end{align}
Note that, in the term $\nu k^2 + \omega_A^2 \eta k^2 / (\omega^2 + \eta^2 k^4)$ of Eq. \ref{eq: fluid_PV_diffusivity}, which defines the width of the response function in time, 
the resistive and viscous damping rates $\eta k^2$ and $\nu k^2$ should be taken as representing eddy scattering (as for resonance broadening) on small scales. 
Also, notice that the mean magnetic field modifies \textit{both} PV diffusivities, via $\omega_A^2$ contributions. 
Comfortingly, on one hand, we recover the momentum flux of 2D fluid turbulence on a $\beta$-plane when we let the large-scale mean magnetic field vanish ($B_0 = 0$).
On the other hand, when the mean magnetic field is strong enough ($ \omega_{A} \gg \omega_R $), the fluctuations are \textit{Alfv\'enic}. 
In this limit $\omega \sim \omega_A \gg \eta k^2 \gg \nu k^2$ (i.e. magnetic Prandtl number $Pm \ll 1$), we have $D_{fluid} \simeq D_{mag}$ and the vorticity flux vanishes, 
i.e. $\Gamma = 0 + \mathcal{O} \big( (\frac{\eta k^2 - \nu k^2}{ \omega_A})^2   \big)$.
This is the well-known `Alfv\'enization' condition, for which the Reynolds and the Maxwell stress cancel, indicating that the driving of the zonal flow vanishes  in the Alfv\'enized state.
There, the MHD turbulence \textit{plays no role in transporting momentum}. 
\subsection{Transition Line and Critical Damping}
The above results corresponds to the lower-right, strong mean field regime in Figure \ref{TDH_prediction}  of \citet{Tobias2007}. 
We can also explain the physics of \textbf{transition line} seen in \citet{Tobias2007} $\beta$-plane simulations, which has weak mean field and is strongly perturbed by MHD turbulence.
This transition line, set by $B_0^2/\eta$, separates the regimes for which large-scale magnetic fields inhibit the growth of zonal flow from those where zonal flows form.
We propose that the transition occurs when the wave becomes critically damped. 
Guided by the parameters from \citet{Tobias2007}, we focus on the transition regime where the dominant mode is at the Rossby frequency ($\omega \sim \omega_R > \omega_A > \eta k^2 \gg \nu k^2$).
Our goal is to find the dimensionless \textbf{transition parameter} ($\lambda$), which characterizes this transition boundary for a particular case (dimensionless parameters $\eta = 10^{-4}$ and $\beta = 5$) in the \citet{Tobias2007} simulation.
Starting with the linear dispersion relation (Eq. \ref{eq: dispersion}), we decompose the frequency into real and imaginary parts ($\omega = \omega_{re} + i\omega_{im} $), leading to
\begin{equation}
\omega_{re} \sim \frac{1}{2} \left(\omega_R + \sqrt {\omega_R^2 + 4 \omega_A^2}\right),
\end{equation}
and 
\begin{equation}\omega_{im} \sim -\left| \frac{\eta k^2 \left(\omega_R -\sqrt {\omega_R^2 + 4 \omega_A^2}\right)}{2 \sqrt {\omega_R^2 + 4 \omega_A^2}} \right|
\end{equation}
in this parameter space.
The transition parameter $\lambda$ is equal to the damping ratio of an oscillatory system and $\lambda=1$ indicates that the system is critically damped (this occurs when $ \omega_{re} = \omega_{im} $).
Thus,
\begin{equation}
	 \lambda \equiv \left|\frac{\omega_{im}}{\omega_{re}}\right|
	 = \frac{\eta k^2 (\omega_R -\sqrt {\omega_R^2 + 4 \omega_A^2})^2 }{4\omega_A^2  \sqrt {\omega_R^2 + 4 \omega_A^2}}.
	\label{keypara}
\end{equation} 
Equating real and imaginary parts ($\lambda = 1$) of the frequency therefore gives the transition boundary. 
In the limit $\omega_R \gg \omega_A$, the transition parameter reduces to 
\begin{equation}
\lambda \; \approx \; \frac{\eta k^2 \omega_A^2}{\omega_R^3} \; \ll 1 ,
 	\label{lambda_largeB}
\end{equation}
indicating the wave is underdamped (see Figure \ref{TDH_prediction}). 
More details are given in Appendix \ref{Appx: energy-stress of Random C}. 
Our results closely match the transition line from \citet{Tobias2007} (see Figure \ref{TDH_prediction}).

\subsection{Cessation of growth via balance of turbulent transport coefficients}

\begin{figure}[h!]\centering
  	\includegraphics[scale=0.37]{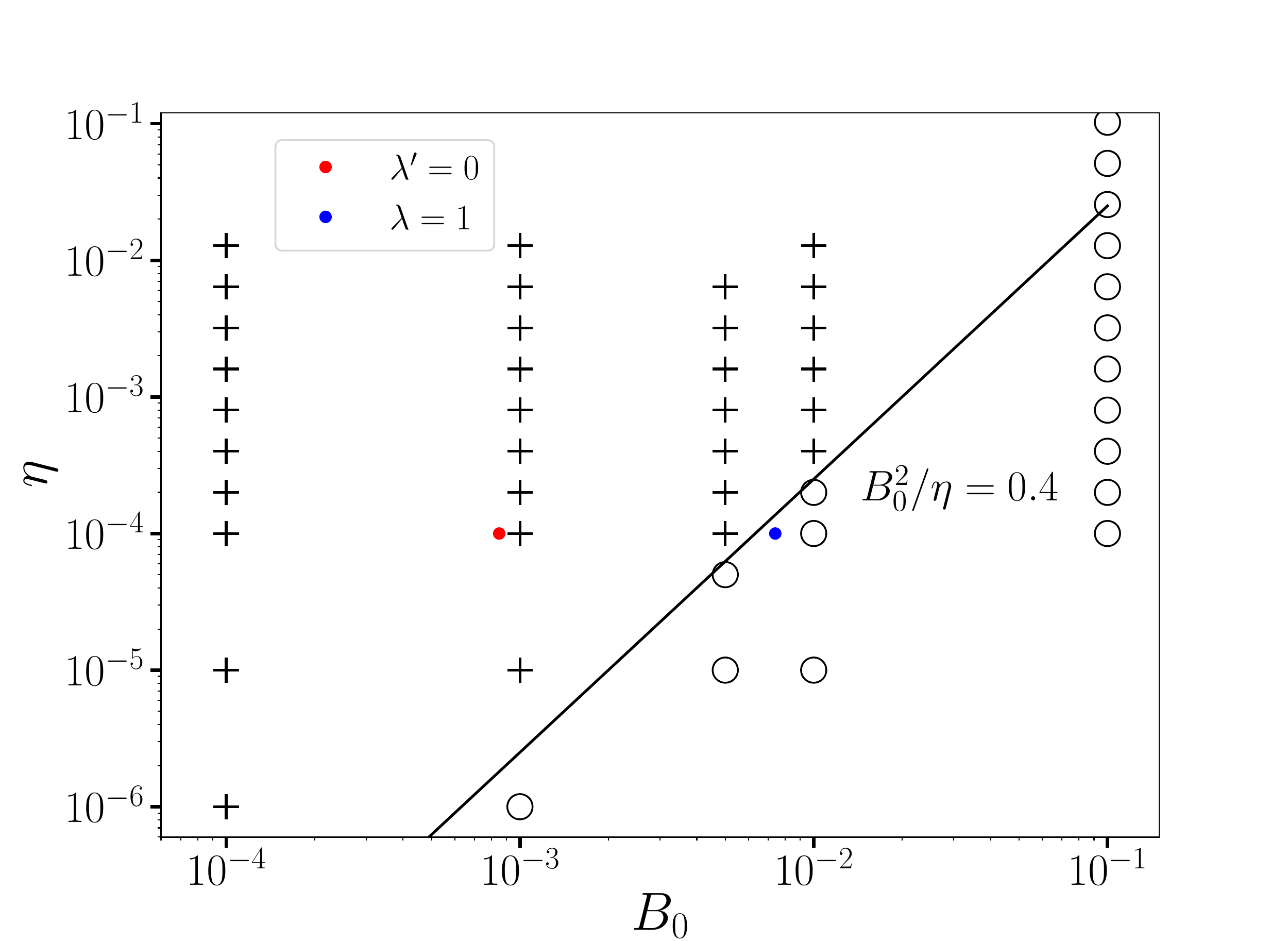}
  	\caption{ Scaling law for the transition between the forward cascades (circles) and inverse cascades (plus signs) from \citet{Tobias2007}. 
	The line is given by $B_0^2 / \eta = 0.4$ .
	The critical transition ($\lambda = 1$) based on our model  predicts a corresponding mean field ($B_0 \sim 7.4 \times 10^{-3}$) in the case of $\eta = 10^{-4}$, labeled by a blue dot. 
	Our result sits close to the transition line in simulation results. 
	The prediction for the critical growth parameter $\lambda^{\prime} = 0 $ is $B_0 \sim 8.4 \times 10^{-4}$, and the associated $B_0$ is an order of magnitude smaller than the one associated with $\lambda =1 $. 
	}
  	\label{TDH_prediction}
\end{figure}

In this section, we are interested in the regime where zonal flow growth ceases. 
To this end, 
we define a \textbf{critical growth parameter} as 
\begin{equation}
    \lambda^{\prime} \equiv \frac{D_{fluid} - D_{mag}}{D_{fluid}}.
    \end{equation}
From this criterion we have:
\begin{equation}	
	\lambda^{\prime} = 1 - \frac{B_0^2 k_x^2}{ \mu_0\rho (\omega^2 +\eta^2 k^4)}.
	\label{lambda^'_largeB}
\end{equation}
When $\lambda^{\prime} =0$, the fluid and magnetic PV diffusivities balance, and the growth of zonal flow vanishes, which is certainly the case for fully Alfv\'enized state (i.e. $\lambda^{\prime} = 0$). 
Figure \ref{TDH_prediction} shows the predicted magnetic field for which $\lambda^{\prime} = 0$ is an order of magnitude smaller than that for $\lambda = 1$.
This is because Rossby-Alfv\'en waves still survive as an underdamped Alfv\'en wave after the growth of zonal flows is turned off.
When $\lambda^{\prime} > 0$, the Maxwell stress is not strong enough to balance the Reynolds stress, so zonal flows are still driven by the Reynolds force. 

\subsection{Comparison of theory with numerical calculations}

In order to assess the validity of our theory, we compare our analysis with results derived from numerical experiments of driven, magnetized turbulence on a doubly periodic $\beta$-plane. 
The numerical results form a small subSection of a much larger unpublished study originally performed by \citet{thd}. The set-up of the model is the same as that described in \citet{Tobias2007}. Namely, we consider a $\beta$-plane in a domain $ 0 \le x,\, y \le 2 \pi$ using pseudospectral methods \citep[see e.g.][]{tobiascattaneo2008}. 

We achieve a steady state of magnetized turbulence by driving the vorticity equation (with $\nu=10^{-5}$) with a small-scale forcing in a band of horizontal wavenumbers $15 \le k_x, \,k_y \le 20$. 
The simulations are started from rest and a small-scale flow is driven initially. 
Eventually, if the magnetic field is weak enough, correlations in the small-scale flow begin to drive a zonal flow via a zonostrophic instability \citep{sy2012}. 
In this case, as time progresses, the zonal flows may grow and merge until a statistically steady state is achieved, with the number of zonal-flow jets depending on the Rhines Scale \citep{rhines_1975, Diamond_2005} and the Zonostrophy Parameter \citep{gsm2008,tm2013}. 
Indeed the final state of  zonostrophic turbulence on a $\beta$-plane (including the number and strength of the jets) may be sensitive to the precise initial conditions. 
The hysteresis may occur between states; see \citet{mct2016}.

If the magnetic field is large enough, then the zonostrophic instability switches off, as shown numerically on a $\beta-plane$ \citep{Tobias2007,dg2016} and on a spherical surface \citep{tdm2011}. 
Theoretically, this suppression of the zonostrophic instability has been described via a straightforward application of QL theory \citep{tdm2011,cp2018}, though as we will show here, this approach does not capture the relevant physics.

\begin{figure}[h!]\centering
  	\includegraphics[scale=0.39]{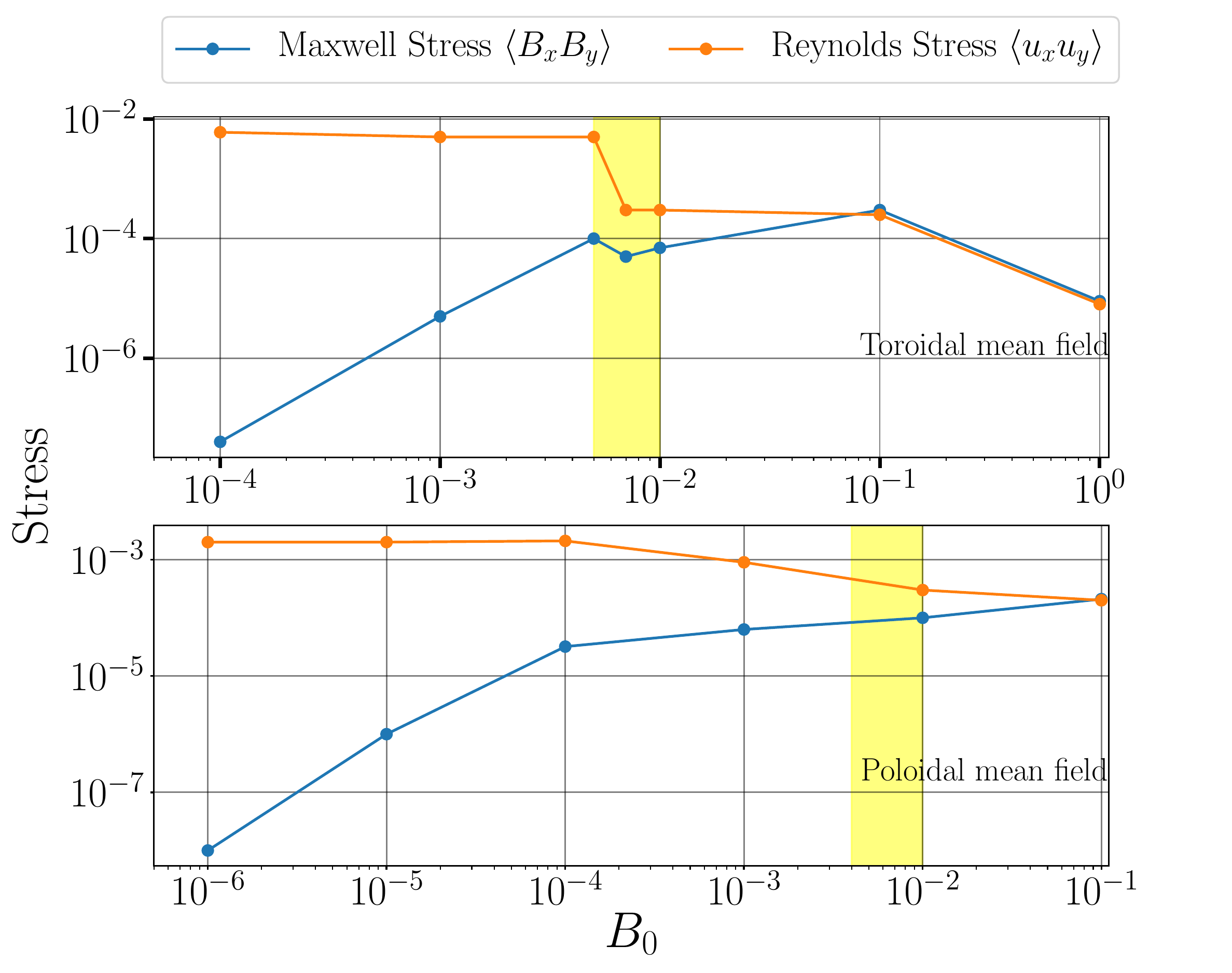}
  	\caption{Average Reynolds stresses (orange line) and Maxwell stresses (blue line) for $\beta$ = 5, $\eta = 10^{-4} $. 
	For the cases of toroidal and poloidal mean field, fully Alfv\'enization happens when $B_0$ intensity is larger than $B_0 = 10^{-1}$ and $B_0 = 6 \times 10^{-2}$, respectively. 
	The yellow-shaded area is where zonal flows cease to grow, following our prediction of the transition parameter $\lambda=1$.
	This is where the random-field suppression on the growth of zonal flow becomes noticeable. 
	}
  	\label{fig:TDH_max-rey}
\end{figure}

Hydrodynamically, for the parameters compared with the theory here $\beta=5$, $\nu=10^{-5}$, $\eta=10^{-5}$, the final state shows the coexistence of turbulence with strong jets on the scale of the computational domain. 
As the magnetic field (either toroidal or poloidal) is increased, eventually it becomes significant enough to switch off the driving of the zonal jets. A simplistic argument would put this down to the magnetic energy  of the small-scale magnetic field (and hence the resultant Maxwell stresses opposing the formation of jets) becoming comparable with the Reynolds stresses that drive the jets. 
However, Figure \ref{fig:TDH_max-rey} shows the Reynolds and Maxwell stresses versus the large-scale field,  and demarcates where the waves are critically damped ($\lambda = 1$).
One sees that the Reynolds stress \textit{already} drops by an order of magnitude, even though $B_0$ is not strong enough to Alfv\'enize the system. 
\textit{This indicates that suppression of zonal flow occurs at values of $B_0$ below the Alfv\'enzation limit}.  
We conjecture this suppression is due to the influence of magnetic fields on the cross-phase in the PV flux (i.e. the Reynolds stress).
We now develop a physical but systematic model of PV mixing in a strongly tangled magnetic field.

\section{PV Mixing in a Tangled Magnetic Field --- beyond QL theory} \label{sec: TangledModel}
\begin{figure}[h!]
\centering
  \includegraphics[scale=0.25]{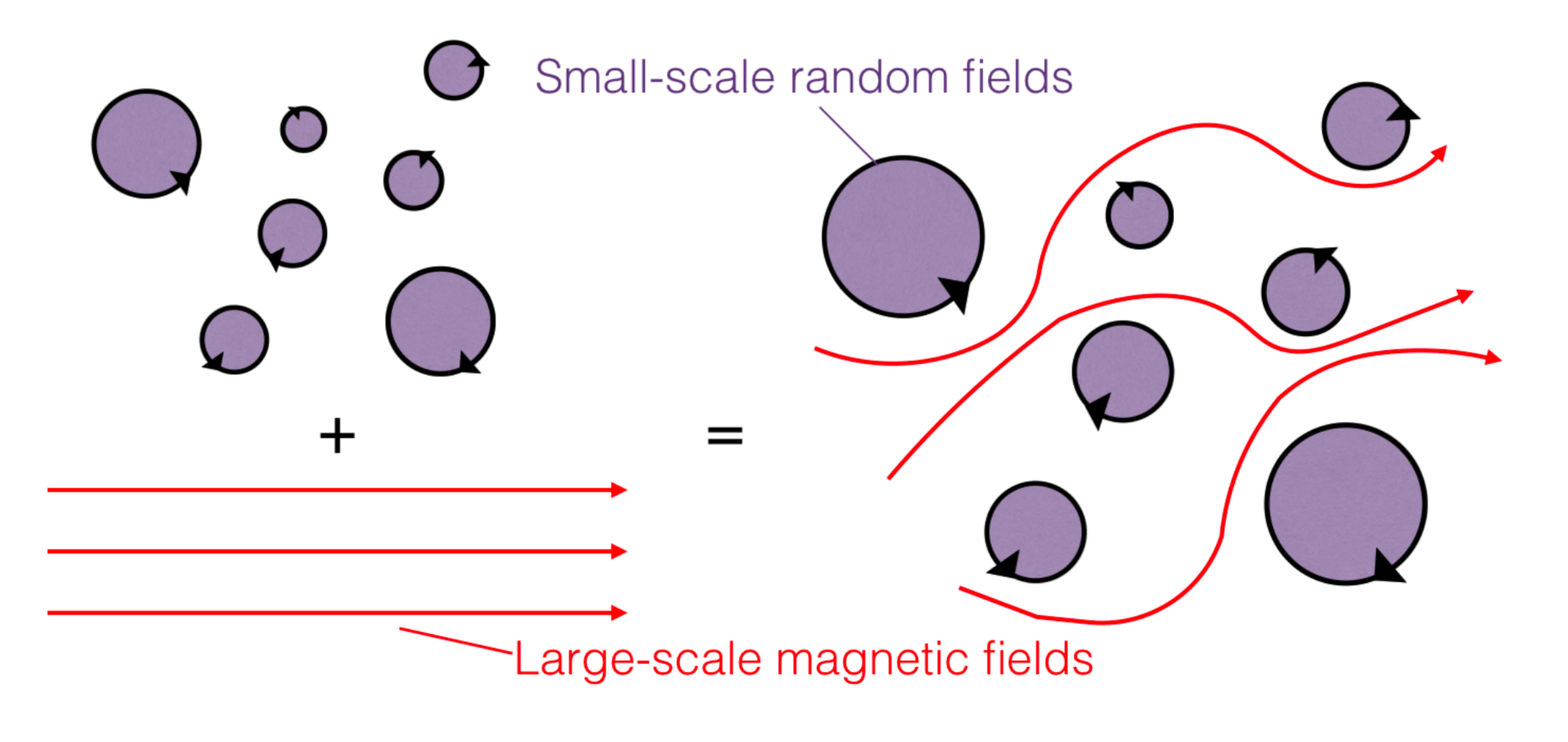}
  \caption{The large-scale magnetic field is distorted by the small-scale fields.
  The system is the `soup' of cells threaded by sinews of open field lines. }
  \label{fig:channel}
\end{figure}

Between the two extremes of the mean field discussed in section \ref{sec:APS_Results}, we are interested in the case of the solar tachocline, where the stretching of mean field by Rossby wave turbulence generates $\widetilde{B}$ and the large-scale magnetic field is not strong enough to Alfv\'enize the system, but remains nonnegligible (i.e. $|\widetilde{B}^2| > B_0^2$ but $ B_0 \neq 0$). 
In this case, the large-scale field lines of the near-constant field will be strongly perturbed by turbulence.
Thus, the magnetic Kubo number is large, for any finite autocorrelation length.
Understanding the physics here requires a model beyond simple QL theory. 
Here we develop a new, nonperturbative approach that we term an  `effective medium' approach.  
\citet{Zel1983_mag_percolation} gave a physical picture of the effect of magnetic fields with $|\widetilde{B}^2| \gg B_0^2$. 
He interpreted the `whole' strongly perturbed problem as consisting of a random mix of two components: a weak, constant field and a random ensemble of magnetic cells, for which the lines are closed loops ($\nabla \cdot \bold{B} = 0$).
Assembling these two parts gives a field configuration of randomly distributed cells, 
threaded by sinews of open lines (see figure \ref{fig:channel}). 
Wave energy can propagate along the open sinews and will radiate to a large distance if the open lines form long-range connections. 
As noted above, this system with strong stochastic fields cannot be described by the simple linear responses retaining $B_0$ only, since $|\widetilde{B}^2| \gg B_0^2$.  

Thus, a `frontal assault' on calculating PV transport in an ensemble of tangled magnetic fields is a daunting task. 
Facing a similar task, \citet{Rosenbluth1977} suggested replacing the 
`full' problem with one where waves, instabilities, and transport are studied in the presence of an ensemble of prescribed, static, stochastic fields. 
Inspired by this idea, we replace the full model with one where PV mixing occurs in an ensemble of stochastic fields that need not be weak --- i.e. $|\widetilde{B}^2|/B_0^2 >1 $ allowed. 
This is accomplished by taking the small-scale fields as spatially uncorrelated ($l_{ac} \rightarrow 0$), i.e. with spatial coherence small.
In simple terms, we replace the `full' problem with one in which stochastic fields are static and uncorrelated, though possibly strong.   
\textit{This way, the magnetic Kubo number remains small---$Ku_{mag}  = l_{ac} |\widetilde{ B} | / (\Delta_{\perp} B_0) <1 $ --- even though $|\widetilde{B}^2 |   \gg B_0^2$}.  
By employing this ansatz, calculation of PV transport in the presence of stochasticity for an ensemble of Rossby waves is accessible to a mean field approach, even in the large perturbation limit. 
Based on this idea, we uncover several new effects including the crucial role of the small-tangled-field ($B_{st}$) in the modification of the cross-phase in the PV flux and a novel drag mechanism that damps flows.
Together, these regulate the transport of mean PV \citep{Kadomtsev1979}.  
We stress again that these effects are not apparent from simple QL calculations.

\subsection{The Tangled Field Model} \label{sec: The Model}

We approach the problem with strongly perturbed magnetic fields ($|\widetilde{B}^2| \gg B_0^2$) by considering an environment with stochastic fields ($B_{st}$) coexisting with an ordered mean toroidal field ($B_0$) of variable strength. 
Notations are listed in Table \ref{tab: notation}. 
The mean toroidal field is uniformly distributed on the $\beta$-plane, while the stochastic component is a set of prescribed, small-scale fields taken as static. 
These small-scale magnetic fields are randomly distributed, and the amplitudes are distributed statistically.

We order the magnetic fields and currents by spatial scales as:
\begin{eqnarray}
	\text{potential field} \;  \;  \;  \;  
	\bold{A} &=  \bold{A_0} + \bold{\widetilde{A}} +\bold{A_{st}} \nonumber \\
	\text{magnetic field}  \; \;  \; \; 
	\bold{B} &=  \bold{B_0} + \bold{\widetilde{B}} + \bold{B_{st}} \nonumber  \\
	\text{magnetic current}  \; \;  \; \; 
	\bold{J} &=  \bold{0} + \bold{\widetilde{J}} + \bold{J_{st}} ,
\end{eqnarray}
where $\bold{J_0} = 0$ for $\bold{B_0}$ is a constant.

\setcounter{table}{1}
\begin{table}[]
\renewcommand{\thetable}{\arabic{table}}
\centering
\caption{Notation} \label{tab: notation}
\begin{tabular}{P{0.16\textwidth} P{0.12\textwidth} P{0.12\textwidth} }
\tablewidth{0pt}
\hline
\hline
Scale  & Magnetic potential field $A$ & Vorticity $\zeta$ \\
\hline
\decimals
Zonal Flow scale  &   $\langle A \rangle \equiv A_0 $ &  $\langle \zeta \rangle$   \\
Wave Perturbation  &$\widetilde{A}$ & $ \widetilde{\zeta}$ \\
Random field average & $\overline{A}$ &  $\overline{\zeta}$ \\
Stochastic field & $A_{st}$ &   \\
\hline
\end{tabular}
\end{table}

\begin{figure}[h!]\centering
\includegraphics[scale=0.14]{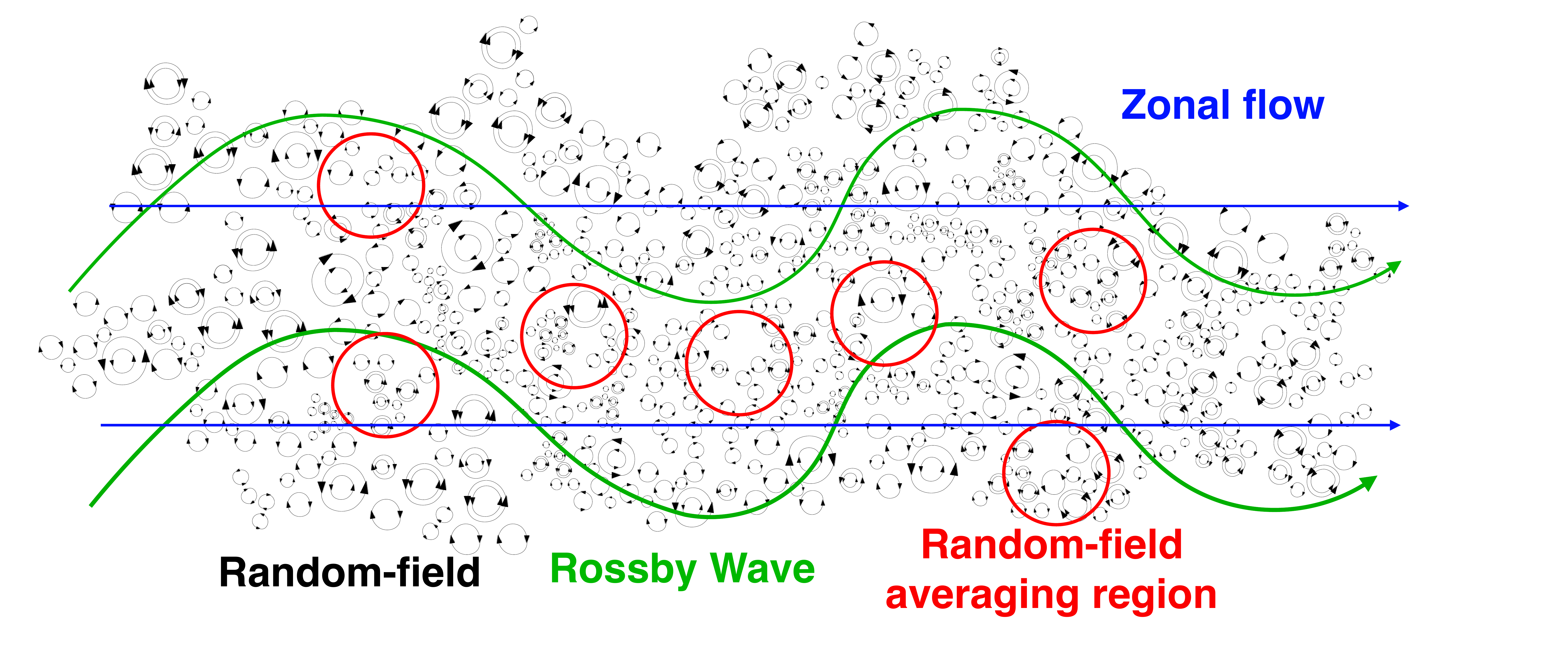}
  	\caption{Length scale ordering. 
	The smallest length scale is that of the random field ($l_{st}$), which looks like a soup of Spaghetti-O's. The random-field averaging region is larger than the length scale of random fields but smaller than that of the Rossby waves. 
	}
  	\label{fig: avg_length scale}
\end{figure}

The waves are described hydrodynamically by:
\begin{eqnarray}
	\text{stream function } \; \; 
	 \bold{\psi} &= \langle \bold{\psi}\rangle + \bold{\widetilde{\psi}}\nonumber  \\
	\text{flow velocity }  \; \;
	\bold{u} &= \langle\bold{u} \rangle + \bold{\widetilde{u}} \nonumber  \\
	\text{vorticity}  \; \; 
	\bold{\zeta} &= \langle\bold{\zeta} \rangle+ \bold{\widetilde{\zeta}},
\end{eqnarray}
where, as before, the $\langle \;\;\rangle$ is an average over the zonal scales ($1/k_{zonal}$) and fast timescales.
For the ordering of wavenumbers of stochastic fields $k_{st}$, Rossby turbulence $k_{Rossby}$, and zonal flows $k_{zonal}$, respectively, we take the scale of spatial average larger than that of Rossby waves.
A length scale cartoon is given in figure \ref{fig: avg_length scale}.

\begin{figure}[h!]
\centering
  \includegraphics[scale=0.13]{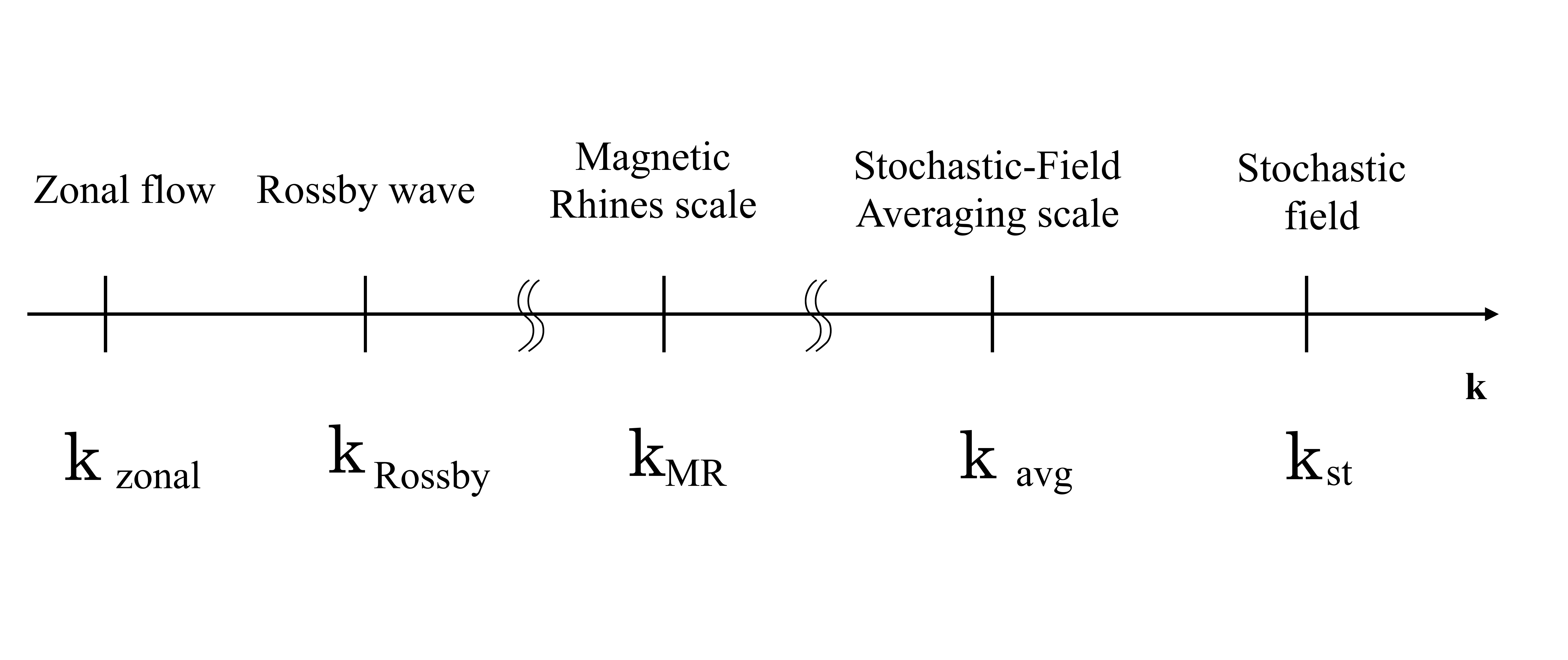}
  \caption{Multi-scale Ordering. The Magnetic Rhines scale separates the regimes of large- and small-length scale. 
   MHD turbulences dominate the system on a smaller length scale and is comprised of Alf\'evn waves and eddies. 
   In this regime, wavenumbers $k$ from high to low are ordered as $k_{st} > k_{avg}$.
   On a larger length scale, however, Rossby waves dominate.
    Here, the scale ordering from high to low wavenumber is: $k_{Rossby} > k_{zonal} $. 
  }
  \label{fig:scale_kspace}
\end{figure}

A procedure to calculate the mean effect of the stochastic fields is to average over the random field, within a window of length scale ($1/|k_{avg}|$):
\begin{equation}
	\bar{F} = \int dR^2 \int d B_{st}\cdot  P_{(B_{st, x},B_{st, y})} F.
\end{equation}
Here, $P_{(B_{st, \, i})}$ is the probability distribution function for the random field, $F$ is the arbitrary function being averaged, and $dR^2$ refers to integration over a region containing random fields. 
This averaging region is larger than the scale of the stochastic field but smaller than the \textbf{Magnetic Rhines scale} \citep[$ l_{MR} \equiv \sqrt{v_{A,\, st}/\beta}$, where $v_{A,\, st}$ is the Alfv\'en velocity with stochastic small-scale fields, and is defined as $v_{A,st} \equiv \sqrt{\overline{B_{st}^2}/\mu_0 \rho}$. See ][]{zel1957, vallis1993} and the Rossby scale.
Thus, we have $ k_{st} >  k_{avg} \gtrsim k_{MR} \gtrsim k_{Rossby} > k_{zonal}$ \citep[see figure \ref{fig:scale_kspace};][]{Tobias2007}.
With this \textbf{random-field average method}, we smooth the effect of small-scale random fields, and so can consider mean field effects of this stochastic system 
(with the assumption that small-scale magnetic fluctuations are spatially uncorrelated).
In this way, the method maintains $Ku <1$ for $| B_{st} |/B_0 >1$ by taking $l_{ac} \rightarrow 0$.
It thus affords us a glimpse of the strong (but random) field regime. 

The novelty and utility of the random-field average method is that it allows the replacement of  the total field due to MHD turbulence (which is difficult to calculate) by moments of the distribution of a static, stochastic magnetic field, which \textit{can} be calculated.  
This is based on the tacit assumption that the perturbation in magnetic fields on the Rossby scale has a negligible effect on the structure of the imposed random fields and its stress-energy tensor.
Put simply, $\overline{(B_{tot})^2 \;} \simeq \overline{B_{st}^2}$ (i.e. first order correction term vanishes, upon averaging), where $\overline{B_{tot}}$ is the averaged total field, regulated by Rossby waves. 

Thus, averaging over the random fields simplifies the analytical model, and we can treat the collective effects of the tangled magnetic field without loss of generality. 
We note here that in the random-field average method, the large-scale field remains the same after averaging ($\overline{B_{0,\, x}}= B_{0, \,x}$). This is because  the mean field is on the zonal length scale, which is larger than the average length scale ($l_{zonal} > l_{avg}$). Moreover, the averaged random field in a selected region $dR^2$ is zero ($\overline{B_{st, \,i}}= 0$, $i = x, \, y$), since the length scale of the stochastic fields is smaller than that of the averaging scale ($l_{st} < l_{avg}$).
Finally, since we assume random fields are spatially uncorrelated ($l_{ac} \rightarrow 0$), we have zero correlation after averaging in $x$- and $y$-direction $\overline{B_{st, \, x} B_{st,\, y}}= 0$ (see Appendix \ref{Appx: energy-stress of Random B}). 

\subsection{Analysis and Results from Tangled Field Model}	\label{sec:ana&Results}
We apply random-field averaging to the vorticity equation first, so as to deal with the nonlinear magnetic term. This yields
\begin{equation}
	\frac{\partial}{\partial t}  \overline{\zeta }
	-\beta\frac{\partial\overline{\psi}}{\partial x} 
	= -\frac{\overline{(\bold{B} \cdot \nabla)\nabla^2 A} }{\mu_0 \rho} 
		+ \nu \nabla^2 \overline{\zeta}.
	\label{eq: avg_vorticity}
\end{equation}
However, we don't apply the random-field average to the induction equation at this stage, as $A_{st}$ is static so that the induction equation for stochastic fields reduces to 
\begin{equation}
\nabla^2 A_{st} = \frac{-1}{\eta} (\bold{B_{st}}  \cdot\nabla)\psi.
 \label{eq:ori_induc}
\end{equation} 
We hold, nonetheless, the general induction equation for mean field, such that $\frac{\partial}{\partial t}   A_0 
	=\bold{B}_0 \cdot \nabla \psi + \eta \nabla^2 A_0.$
Combining Eq. \ref{eq: avg_vorticity} and \ref{eq:ori_induc}, we have 
\begin{equation}
	\frac{\partial}{\partial t}  \overline{\zeta }
	-\beta\frac{\partial \overline{\psi}}{\partial x} 
	= 
		\frac{1}  
		{ \eta \mu_0 \rho} 
		\frac{\partial}{\partial y} (\overline{B_{st,\, y}^2} \frac{\partial}{\partial y} \overline{\psi})  
		-\dfrac{B_0 }{\mu_0 \rho} 
		\frac{\partial (\nabla^2 \overline{A}_0)}{\partial x} 
		+ \nu \nabla^2 \overline{\zeta}.
		\label{eq:rad_vorti}
\end{equation}
Next, we consider the vorticity wave perturbation after applying the random-field average:
\begin{equation}
	\frac{\partial}{\partial t}  \widetilde{\zeta} 
	+\beta \widetilde{u}_y
	+ \widetilde{u}_y \frac{\partial}{\partial y} \overline{\zeta}
	=  
	\frac{\partial}{\partial y} \frac{\overline{B_{st, \,y}^2} \frac{\partial}{\partial y} \widetilde{\psi} }{\eta \rho\mu_0 }
	-\dfrac{B_0}{\mu_0 \rho} 
		 \frac{\partial  (\nabla^2 \widetilde{A}_0)}{\partial x}
	+\nu  \nabla^2 \widetilde{\zeta}.
\label{eq:ran_vorti_perturb}
\end{equation}
Equation \ref{eq:ran_vorti_perturb} is formally linear in pertubations and allows us to calculate the response of the vorticity in the presence of tangled fields, namely 
\begin{equation}
	\widetilde{\zeta_k} 
	=
	\bigg(\frac{-i}{ \omega
				+i\nu k^2 
				+ \frac{ i \overline{B_{st,\, y}^2} k _y^2 }{\mu_0 \rho  \eta k^2} 
				+  \frac{-B_0^2 k _x^2 }{\mu_0 \rho (\omega + i\eta k^2 )} 
				}\bigg)  								
				\widetilde{u}_{y,\, k}\big(
				\frac{\partial}{\partial y} \overline{\zeta} + \beta 
				\big).
	\label{eq:ran_response of vorti}
\end{equation}
The effective medium Rossby-Alfv\'en dispersion relation can be derived from this Eq. \ref{eq:ran_response of vorti}, and is given by
\begin{equation}
	 \bigg(  
	 \omega - \omega_R 
	 + \frac{i \overline{B_{st,\, y}^2} k_y^2 }{\mu_0 \rho \eta k^2}
	 + i\nu k^2 
	 \bigg) 
	 \bigg(\omega +  i \eta k^2  \bigg) 
	 = 
	 \frac{B_0^2 k_x^2 }{\mu_0 \rho}.
\end{equation}
With $B_{st} = 0$, we recover the standard Rossby-Alfv\'en waves described in Section \ref{sec:APS_Results}.
Now, the average over the zonal scales and the assumption that zonal flows are still noticeable ($ \frac{\partial}{\partial x} \langle \; \rangle \rightarrow 0$) give us the mean, `double-averaged', vorticity equation:
\begin{equation}
	\frac{\partial}{\partial t}  \langle \overline{\zeta} \rangle
	= - \frac{\partial}{\partial y} \langle
	 \overline{\Gamma} 
	 \rangle
	+ \frac{1}{ \eta\mu_0\rho}  \frac{\partial}{\partial y}
	\bigg( 
	\langle \overline{B_{st, \, y}^2} \rangle \frac{\partial}{\partial y} \langle\overline{\psi}
	\rangle
	\bigg)
	+\nu  \nabla^2 \langle\overline{\zeta} \rangle,
	\label{random_evl_vorticity}
\end{equation}
where term $\overline{\Gamma}$ here is the mean PV flux such that
$
       \langle \overline{\Gamma}  \rangle = \langle \widetilde{u}_y\widetilde{\zeta}\rangle	
       \equiv -D_{PV} \big(\frac{\partial}{\partial y} \langle \overline{\zeta}\rangle + \beta \big).
$
Integrating equation (\ref{random_evl_vorticity}) in $y$ yields
\begin{equation}
\frac{\partial}{\partial t}  \langle u_x \rangle
	=  \langle
	 \overline{\Gamma} 
	 \rangle
	- \frac{1}{ \eta\mu_0\rho}  
	\langle \overline{B_{st,\, y}^2} \rangle  \langle u_x\rangle + \nu \nabla^2 \langle u_x 
	\rangle.
	\label{eq:zonal_drag}
\end{equation}
In addition to the mean PV flux, note the drag term $\frac{1}{ \eta\mu_0\rho}  
	\langle \overline{B_{st, \, y}^2} \rangle  \langle u_x
	\rangle$ that results from the $\langle \mathbf{J_{st}} \times \mathbf{B_{st}} \rangle$ force.
The mean-square random field effect $\langle \overline{B_{st, \, y}^2} \rangle$ appears both in the mean flux $\langle \overline{\Gamma} \rangle$ and in the drag.

We now discuss both effects. 
First, the mean PV flux $\langle \overline{\Gamma} \rangle$ is affected by both large- and small-scale fields. 
The mean PV flux as a function of both large-scale mean field  $B_0$ and the mean-square stochastic field $\overline{B_{st,\, y}^2}$ may be expressed as: 
\begin{equation}
	\overline{\Gamma} 
	=  -\sum_{k} |\widetilde{u}_{y,\, k} |^2 
	C_k
		\big(\frac{\partial}{\partial y} \overline{\zeta} + \beta \big),
		\label{eq: random_vorti_flux}
\end{equation}
where the resonance function (phase coherence) $C_k$, which defines the effective decorrelation time $\tau_{c, \, k}$, is:
\begin{equation}
	C_k \equiv 
	\frac{\nu k^2 
		+\frac{ \omega_A^2\eta k^2} {\omega^2 + \eta^2 k^4} 
		+  \frac{ \overline{B_{st, \, y}^2} k _y^2 }{\mu_0 \rho  \eta k^2} 
		}
		{\omega^2\big( 1- \frac{\omega_A^2 }{\omega^2 + \eta^2 k^4} \big)^2	
		+\big(
		\nu k^2 
		+ \frac{\omega_A^2 \eta k^2}{\omega^2  + \eta^2 k^4} 
		+ \frac{ \overline{B_{st, \, y}^2} k _y^2 }{\mu_0 \rho  \eta k^2}
		\big)^2
		}.
	\label{eq: PV flux constant}
\end{equation}
Observe that both $\overline{B_{st,y}^2 }$ and $B_0^2$ tend to reduce $\overline{\Gamma}$ for a fixed level $\langle \widetilde{u}_y^2 \rangle$.
Compared with Eq. \ref{eq: mag_PV_diffusivity}, an additional term due to the mean-square stochastic field $ \overline{B_{st, \, y}^2} k _y^2 /\mu_0 \rho  \eta k^2$ plays a role in the cross-phase by modulating the prefactor $C_k$ that enters the PV diffusivity. 
The scaling indicates that the zonal flow can be suppressed by the stochastic field effect in the cross-phase. 
Moreover, when the mean field is weak ($B_0^2 \ll \overline{B_{st}^2}$), the cross-phase effect is dominant.
\textit{This is consistent with the observed drop of the Reynolds stress when the mean field is weak} (see figure \ref{fig:TDH_max-rey}) 

Note that if we turn off the large-scale magnetic field, eddy scattering (resonance broadening) appears both via the turbulent viscosity $\nu k^2$ and via the stochastic field $\overline{B_{st, \, y}^2} k _y^2 $, leading to the modification of the phase coherence $C_k$. 
As stochastic fields become stronger, so does the eddy scattering effect.  
Note that this effect on the PV flux originates via the Reynolds stress and not the Maxwell stress because of our \textit{a priori} postulates of a pre-existing ambient stochastic field and the ansatz $\overline{B_{st, \, x}B_{st, \, y}} = 0$, which lead to zero Maxwell stress by construction.
However, even though the Maxwell stress vanishes, \textit{the mean-square random fields ($\overline{B_{st}^2}$) can still modify the cross-phase of the (fluid) Reynolds stress}.
Thus, we see that large- and small- scale magnetic fields have synergistic effects on the mean PV flux $\overline{\Gamma}$.

Second, the mean-square stochastic fields also set the magnetic drag that modifies the evolution of vorticity, given by Eq.~\ref{eq:zonal_drag}. 
The physics of this drag can be elucidated via an analogy between random fields and a tangled network of springs \citep{montroll1955effect, alexander1981excitation}. 
From the second term in Eq. \ref{eq:zonal_drag}, we can infer a drag constant $\alpha$ ($F_{drag} \propto  - \alpha  \langle u_x \rangle $).
In the absence of rotation ($\beta =0$), one can write down the dispersion relation, and find
\begin{equation}
	\omega^2 
	+ i\underbrace{(\alpha +\eta k^2)}_{\text{drag + dissipation}}\omega 
	-\underbrace{\left(\frac{ \overline{B_{st, \, y}^2} k_y^2}{ \mu_0\rho} + \frac{ B_0^2 k_x^2}{ \mu_0\rho }  \right) }_{\text{effective spring constant}}
	=0,
\end{equation}
where the drag coefficient is $
	\alpha \equiv \overline{B_{st,y}^2} k_y^2  / \mu_0 \rho  \eta k^2.
$
This shows that the effective spring constant is set by the mean field and the stochastic field ($K = (\overline{B_{st, \, y}^2} k_y^2+ B_0^2 k_x^2)/ \mu_0\rho $).
In the typical case where the mean-square stochastic field is dominant, the drag constant can be approximated as $\alpha \sim K / \eta k^2$, i.e. the effective $\textbf{drag force}$ is given by the ratio of the effective elasticity ($K$) to the dissipation ($\eta k^2$). 
This implies the tangled fields and fluids define a \textbf{resisto-elastic} medium \citep{brenig1971theory, kirkpatrick1973, harris1977low}. 
This dissipative character of the medium is due to the fact that for our system $\frac{\partial}{\partial t} A_{st} =0$ (i.e. static stochastic fields), so inductive effects vanish. 

\begin{figure}[h!]\centering
  	\includegraphics[scale=0.20]{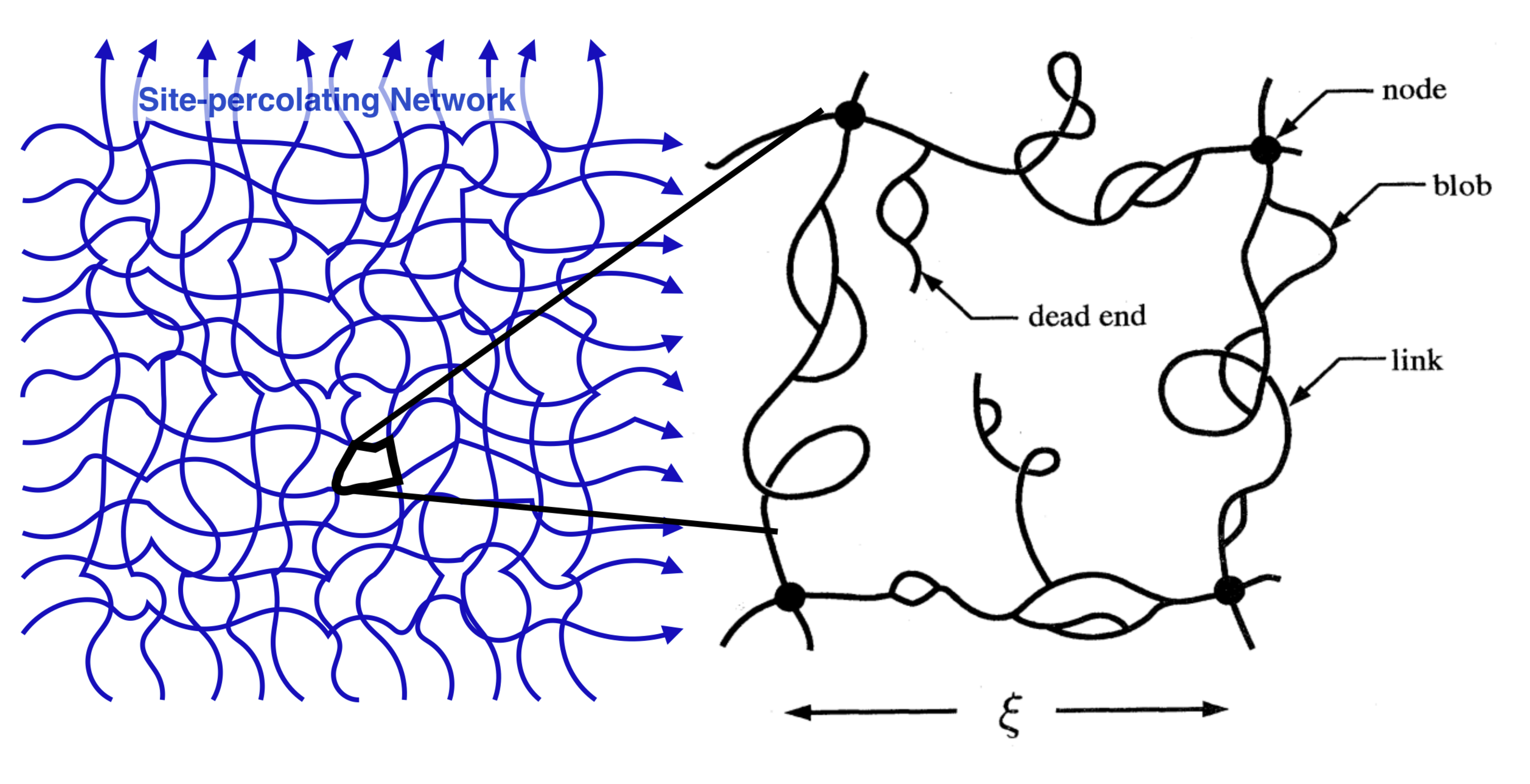}
  	\caption{ Site-Percolation Network.
	Schematic of the nodes-links-blobs model (or SSdG model, see \citealt{Skal1974, de1976relation, naka1994}).
	This depicts the resisto-elastic medium formed by small-scale stochastic fields. 
	}
  	\label{fig:percolation}
\end{figure}

A way to visualize the dynamics of vorticity in this system, dominated by strong stochastic fields, is to again think of the PV as `charge density' $\rho_{PV}$, following from the same understanding discussed in Section  \ref{sec: MHD beta plane}.
The mean vorticity evolution is now given by
\begin{equation}
	  \frac{\partial}{\partial t} \langle  \rho_{PV} \rangle
				=-\frac{\partial}{\partial y}\langle \widetilde{u}_y\widetilde{\rho}_{PV} \rangle
				+  \frac{1}{\eta\mu_0\rho} \frac{\partial}{\partial y}
	\bigg( 
	\langle \overline{B_{st, \, y}^2} \rangle \frac{\partial}{\partial y} \langle\overline{\psi}
	\rangle
	\bigg),
	\label{eq: charge flow through percolation}
\end{equation}
where the second term of Eq. \ref{eq: charge flow through percolation} is obtained from the term $\frac{1}{\rho} \frac{\partial}{\partial y} \langle B_{st} J_{st} \rangle$ and substituting $J_{st} = - B_{st} \nabla \overline{\psi}/\eta$.
This states that in this strong magnetic turbulent case, the charge density is also redistributed by the drag of small-scale stochastic fields, which form a resisto-elastic network (see figure \ref{fig:percolation}). 
Since $\langle \widetilde{B}^2 \rangle / B_0^2 \sim Rm$, the drag due to the small-scale field is larger than that of the mean field.

All in all, mean-square random fields can influence the evolution of zonal flow not only by changing the phase correlation of PV flux but also by changing the structure of the resisto-elastic network.  
As mean-square random fields are magnified, the PV flux drops, while the drag is enhanced.  

\subsection{Transition Parameters for Tangled Fields}

Following the same logic as in Section \ref{sec:APS_Results}, we examine the growth of zonal flow and the properties of wave, under the influence of strong stochastic fields. 
We derive the dimensionless transition parameter $\lambda$, which quantifies the criticality of damped waves. 
A regime where the intensity of the stochastic field is strong enough so that the mean field, resistivity, and viscosity are negligible ($\omega_{R} \sim \omega_{re} >\omega_{st} \gg \eta k^2 \gg \nu k^2 \sim \omega_A^2$) is identified.
For this case, we have: 
$
	\omega_{re} \sim\omega_R
$ and 
$
	\omega_{im} \sim - \omega_{st}^2/\eta k^2,
$
where Alfv\'en frequency ($\omega_{st}$) of collective random fields is defined as $\omega_{st} \equiv \sqrt{ \overline{B_{st, \, y}^2} k_y^2/\mu_0 \rho}$. 
Thus, the \textbf{transition parameter} for this regime is (see Eq. \ref{lambda_largeB}):
\begin{equation}
	\lambda \equiv \left|\frac{\omega_{im}}{\omega_{re}}\right| = \frac{\omega_{st}^2}{\eta k^2 \omega_R} 
	=  \frac{\omega_{st}^2 l_{MR}^2}{\eta k_x \widetilde{u}},
\end{equation}
where $\widetilde{u}$ is the typical eddy velocity and $l_{MR}$ is the magnetic Rhines scale. 
When $\lambda =1$, the wave is critically damped. 

The critical growth parameter ($\lambda^{\prime}$), that defines the growth of zonal flow, is now given by (see Eq. \ref{lambda_largeB}) 
\begin{equation}
	\lambda^{\prime} 
	\equiv
	 \frac{\langle \overline{\Gamma} \rangle - \frac{1}{ \eta\mu_0\rho}  
	\langle \overline{B_{st, \, y}^2} \rangle  \langle u_x
	\rangle 
}{\langle \overline{\Gamma} \rangle }.
\end{equation}
From Eq. \ref{eq:zonal_drag} one should notice that the zonal flow stops growing when the drag force cancels the PV flux ($
\langle \overline{\Gamma} \rangle = \frac{1}{ \eta\mu_0\rho}  
	\langle \overline{B_{st, \, y}^2} \rangle  \langle u_x
	\rangle,
$ ignoring the viscosity).
This corresponds to $\lambda^{\prime} = 0$, where $\langle \overline{\Gamma} \rangle$ is quenched by $\langle \overline{B_{st, \, y}^2} \rangle$. 

Finally, one might ask how this suppression of PV flux relates to the related phenomenon of the quenching of turbulent magnetic resistivity ($\eta_T$) in a weak mean field system \citep{zel1957}. 
The answer can be shown by looking into the PV diffusivity derived from Eq. \ref{eq: random_vorti_flux} in a weak mean field system ($B_0 \rightarrow 0$): 
\begin{equation}
	D_{PV} 
	 = \sum_{k} |\widetilde{u}_{y, \, k} |^2 
	\dfrac{\nu k^2 
		+  \frac{ \overline{B_{st, \, y}^2} k _y^2 }{\mu_0 \rho  \eta k^2} 
		}
		{\omega^2	
		+\bigg(
		\nu k^2 
		+ \frac{ \overline{B_{st, \, y}^2} k _y^2 }{\mu_0 \rho  \eta k^2}
		\bigg)^2
		}.	
\end{equation}
Recall the form of the quenched turbulent resistivity \citep{GruzinovDiamond1994, GruzinovDiamond1996} :
\begin{equation}
 	\eta_T = \sum\limits_{k} |\widetilde{u}_k^2| \frac{ \tau_{c, \, k}}{1+Rm\frac{v_{A, \, 0}^2}{\langle  \widetilde{u}^2 \rangle}}
= 
\sum\limits_{k} |\widetilde{u}_k^2| \frac{ \tau_{c, \, k}}{1+\frac{v_{A,\,st}^2}{\langle  \widetilde{u}^2 \rangle}},
\end{equation} 
where $v_{A,\,st}^2 \equiv \overline{B_{st,\, y}^2} / \mu_0 \rho$ and  $v_{A,\,0}^2 \equiv B_{0}^2 / \mu_0 \rho$.
This is based on the Zel'dovich relation $\overline{B_{st, \, y}^2} \sim  Rm B_0^2
$ \citep{zel1957} in a high magnetic Reynolds number system.  
To compare these two diffusivities $D_{PV} $ and $\eta_T$, one can rewrite the expression of $D_{PV}$ as
\begin{equation}
	D_{PV} = \sum\limits_{k}|\widetilde{u}_{y, \, k}|^2 \frac{ \alpha/\omega^2}{1+( \alpha/\omega^2 )^2}, 
\end{equation}
where $\alpha \equiv  \overline{B_{st, \, y}^2} k _y^2 /\mu_o \rho \eta k^2$ is the effective drag coefficient. 
The term $\alpha/\omega^2$ in the numerator defines the effective decorrelation time $\tau_c$.  
This leads to the inference that both the PV diffusivity and the turbulent magnetic resistivity in a weak magnetic field are reduced by the effect of mean-square random fields $\overline{B_{st, \, y}^2}$. 
Though differences arise from different assumptions about the small-scale magnetic field (for PV, $\widetilde{B}$ is static; for $\eta_T$ the analysis considers dynamic $\widetilde{B}$, see \citealt{Fan2019}), the basic physics of these two quenching effects is fundamentally the same.


\section{Conclusion} \label{sec: discussion and conclusion}

In this paper, we have developed and elucidated the theory of PV mixing and zonal flow generation, for models of Rossby--Alfv\'en turbulence with two different turbulence intensities. 
Our most novel model considered the large fluctuation regime ($\langle \widetilde{B}^2 \rangle/B_0^2  > 1 $) --- where the field is tangled, not ordered.
For this, we developed a theory of PV mixing in a static, stochastic magnetic field. It is striking that 
this model problem is amenable to rigorous, systematic analysis yet yields novel insights into the broader questions asked. 

Our main results can be summarized as follows. 
First, we have defined the \textit{magnetic Kubo number} and demonstrated the importance of ensuring  $Ku \ll 1$ for the application of QL theory to a turbulent magnetized fluid. 
In this regime, we have derived the relevant QL model for turbulent transport and production of jets and shown the utility of the \textit{critical damping parameter} in determining the transition between jet drive and suppression by the magnetized turbulence.

A striking result is that numerical  experiments show how magnetic fields may significantly reduce the Reynolds stresses, which drives jets, well before the critical mean field strength needed to bring the Maxwell and Reynolds stresses into balance, i.e. before Alfv\'enization. 
This is important and demonstrates that the magnetic field acts in a subtle way to change the transport properties --- indeed, even more subtle than was previously envisaged. 
The explanation of this effect required the development of a new model of PV mixing in a tangled, disordered magnetic field. This tractable model has $Ku_{mag} < 1$, because the tangled field is delta correlated and allows the consideration of strong stochastic fields $\overline{B_{st}^2}/B_0^2 >1$. We use a `double average' procedure over random-field scales and mesoscales that allows treatment of the wave and flow dynamics in an effective resistive-elastic medium.

We identify two principle effects as the crucial findings:
	\begin{enumerate}
	\item A modification (reduction) of the cross-phase in the PV flux by the \textit{mean-square} field $\overline{B_{st}^2} $.
	This is in addition to $\omega_A^2$ effects, proportional to $B_0^2$, which appears in QL theory. 
	Note that this is not a fluctuation quench effect. 
	\item A magnetic drag, which is proportional to $\langle \overline{B_{st}^2} \rangle$, on the mean zonal flow.
	The scaling of $ \langle \overline{B_{st}^2} \rangle / \eta$ resembles that of the familiar magnetic drag in the `electrostatic' limit, with $\overline{B_{st}^2}$ replacing $B_0^2$.
	Note that the appearance of such a drag is not surprising, as stochastic fields are \textit{static}, so $\frac{\partial}{\partial t} A_{st} \rightarrow 0$. 
	\end{enumerate}

The picture discussed in this paper is analogous to that of dilute polymer flows, in which momentum transport via Reynolds stresses is reduced, at roughly constant turbulence intensity, leading to drag reduction.
The similarity of the Oldroyd-B model of polymeric liquids and MHD is well known \citep{oldroyd1950formulation, oldroyd1951motion, bird1987dynamics, rajagopal1995exact,op2003,bhp2009}. 
A Reynolds stress phase coherence reduction related to mean-square polymer extension is a promising candidate to explain the drag reduction phenomenology. 

More generally, this paper suggests a novel model of transport and mixing in 2D MHD turbulence derived from considering the coupling of turbulent hydrodynamic motion to a fractal elastic network \citep{broadbent1957percolation, rammal1983random, rammal1983nature, rammal1984spectrum, mandelbrot1984physical, ashraff1988density}. 
Both the network connectivity and the elasticity of the network elements can be distributed statistically and can be intermittent and multiscale.
These would introduce a packing fractional factor to $C_k$ in the cross-phase, i.e. $\langle \widetilde{B}^2 \rangle \rightarrow p \langle \widetilde{B}^2 \rangle$ in $C_k$, where $0< p < 1$ is probabilities of sites.
This admittedly crude representation resembles that of the mean field limit for `fractons' \citep{alexander1982density}.
Somewhat more sophisticated might be the form $\langle \widetilde{B}^2 \rangle \rightarrow (p-p_{c})^{\gamma}|\widetilde{B}|^{2\epsilon}$, where $p_c$ is the magnetic activity percolation threshold, and $\gamma$, $\epsilon$ are scaling exponents to be determined \citep{stanley1977cluster}. 
We also speculate that the back-reaction (at high $Rm$) of the small-scale magnetic field on the fluid dynamics may ultimately depend heavily on whether or not the field is above the packing `percolation threshold' for long-range Alfv\'en wave propagation. 
Such long-range propagation would induce radiative damping of fluid energy by Alfv\'enic propagation through the stochastic network. 

We also note that this study has yielded results of use in other contexts, most notably that of magnetized plasma confinement where the field is stochastic, as for a tokamak with resonant magnetic perturbations (RMP).
Indeed, recent experiments \citep{Kriete2019APS,  Neiser_2019, Schmitz_2019} have noted a reduction in shear flow generation in plasmas with RMP. 
This reduction causes an increase in the low/high confinement regime power threshold. 

Finally, in the specific context of modeling tachocline formation and dynamics, this analysis  yields a tractable model of PV transport, which can incorporate magnetic effects into hydrodynamic models. 
In this paper, we ignore the perturbation of random fields $\widetilde{B}$ (see Appendix \ref{Appx: energy-stress of Random B}).
Here $\widetilde{B}^2$ can be replaced by $\langle \overline{B_{st}^2} \rangle$ and be estimated using the Zel'dovich value $\overline{B_{st}^2} \sim B_0^2 Rm$.
The model suggests that the `burrowing' due to meridional cells that drives tachocline formation will be opposed by relaxation of PV gradients (not shears!) and the resisto-elastic drag.  
The magnetic-intensity-induced phase modification will reduce PV mixing relative to the prediction of pure hydrodynamics. 
Thus, it seems fair to comment that neither the model proposed by \citet{Spiegel1992}, nor that by \citet{gough1998inevitability} is fully ``correct". 
The truth here is still elusive, and `neither pure nor simple' (apologies to Oscar Wilde).

\acknowledgments
The authors thank Steve Tobias for providing the numerical simulation data used in this paper, and also for many useful discussions and a critical reading of the manuscript. We also thank Xiang Fan, \"Ozg\"ur G\"urcan and David Hughes for stimulating discussions on this topic.
The authors also acknowledge participants in the 2018 Chengdu Theory Festival and the 2019 Aix Festival de Th\'eorie.
This research was supported by the US Department of Energy, Office of Science, Office of Fusion Energy Sciences, under award No. DE-FG02-04ER54738.


%



\appendix
\section{Details of QL theory Predictions} \label{Appx: energy-stress of Random C}
\begin{figure}[h!]\centering
  	\includegraphics[scale=0.5]{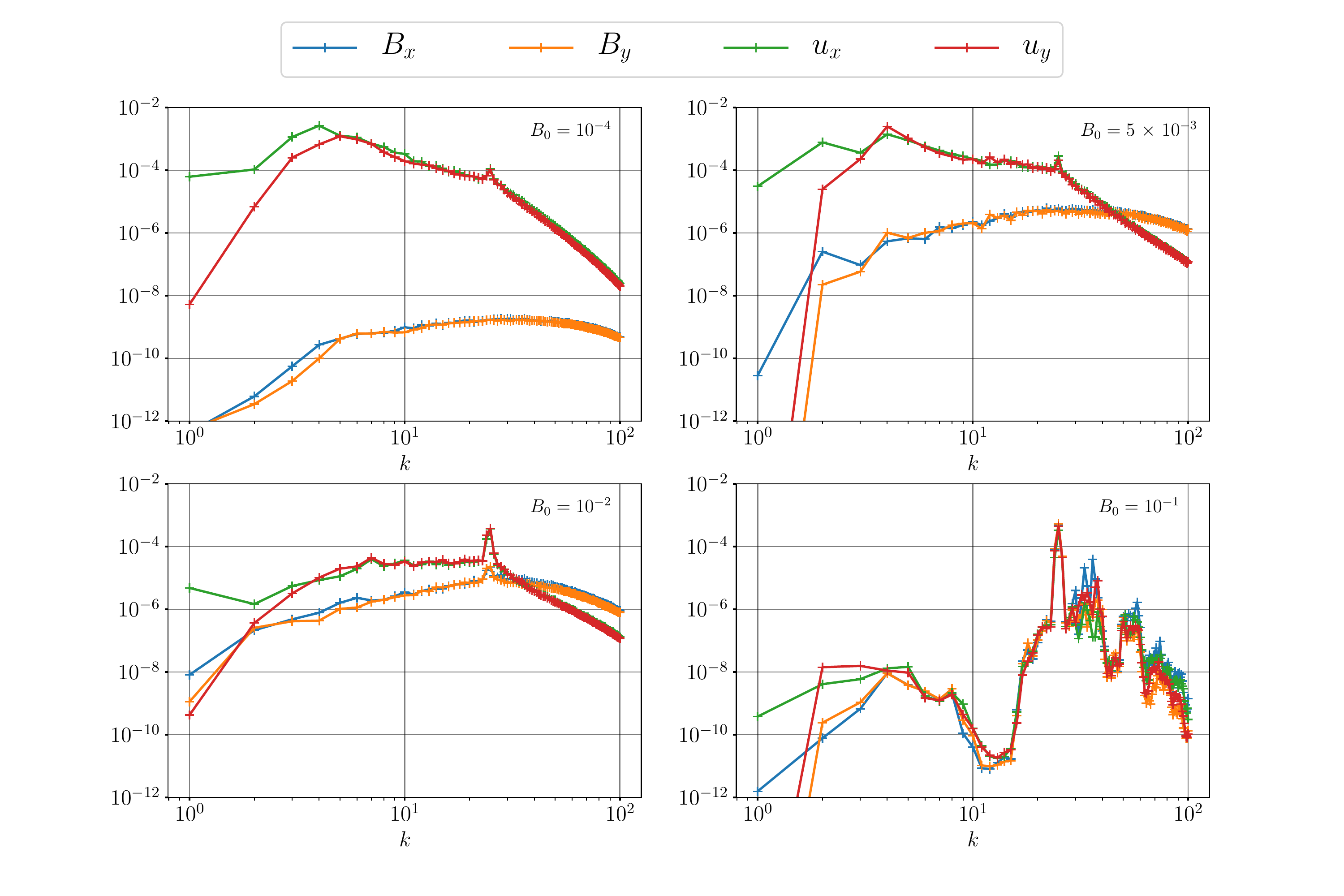}
  	\caption{Spectra for $v_x$, $v_y$, $B_x$ and $B_y$ for an imposed toroidal field with $\langle B_x \rangle \,$(defined as $B_0$)$\, = 10^{-4}$, $5\times10^{-3}$, 						
	$10^{-2}$ and $10^{-1}$ for $\beta = 5$ and $\eta = 10^{-4}$. 
	}
  	\label{TDH_spectra}
\end{figure}

Here we investigate the corresponding prediction of transition parameter $\lambda = 1$ from Eq. (\ref{keypara}) and compare it to the transition line in \citet{Tobias2007}.
First, we find $\lambda =  2.87\times 10^{-10}$ and $8.04$ for $B_0 = 5\times10^{-3}$ and $B_0 = 1\times10^{-2}$, respectively.
These spectra of velocities and field are from the present ongoing paper (Tobias, Diamond, and Hughes et al.).
The peak of wavenumber in the spectra from the top left to the bottom right is $k  \sim 3.6$, $4$, $23.5$, and $25.5$ (see Figure \ref{TDH_spectra}). 
We obtain four transition parameters for these four spectra with different mean field $B_0$, and find that the transition ($\lambda = 1$) occurs when $5 \times 10^{-3} < B_0 < 1\times 10^{-2}$. 
The corresponding regime of magnetic intensity for the occurrence of the transition is shaded yellow (see Figure \ref{fig:TDH_max-rey}).
We also plot $k_x$ vs. $B_0$ and assume that the wavenumber is a linear function of $B_0$, and hence the prediction of transition is narrowed down to magnetic toroidal field $B_0 \sim 7.4 \times 10^{-3}$. 
\textit{This result is consistent with the simulation from \citet{Tobias2007} }(see Figure \ref{TDH_prediction}).
Similarly, if we check the \textbf{critical growth parameter} $\lambda^{\prime} = 0$ with the same method, we would find out that the zonal flow stops growing at $B_0 \sim 8.5 \times 10^{-4}$, which is at magnitude of an order lower than $\lambda = 1$.
\section{Collective Random Magnetic Fields} \label{Appx: energy-stress of Random B} 

We check the validity of the assumption for ignoring changes in random fields on the small, stochastic scales ($l_{st}$) due to Rossby wave straining, after applying the \textbf{random-field average method}. 
Here we turn off the mean field ($B_0 = 0$) and consider the random fields only ($B_{tot} = 0 + \widetilde{B} +B_{st} $). 
As the Rossby wave may perturb the small-scale random field, we can write the total magnetic field as
\begin{equation}
	B_{tot} \equiv B_{st} + \widetilde{B} ,
	\label{eq: A1}
\end{equation}
where $B_{tot}$ is the total random field including the effect of the Rossby turbulence, $B_{st}$ is stochastic fields, and $\widetilde{B}$ is \textit{the change of the magnetic field induced by $B_{st}$}. 
Also, the linear response of collective fields ($\delta \overline{B_{tot}}$) and the random fields ($\delta B_{st}$) have the relation:
\begin{equation}
	\frac{\delta \overline{B_{tot}}}{\overline{B_{tot}}} =\frac{\delta B_{st}}{\widetilde{B}}.
	\label{eq: A2}
\end{equation}
Note that collective fields $\overline{B_{tot}}$ are at Rossby-wave scale ($k_{Rossby}$) after applying the \textbf{random-field average method}.
Combining Eq. \ref{eq: A1} and \ref{eq: A2}, we have
\begin{equation}
	B_{tot} \equiv B_{st} + \frac{\delta B_{st}}{\delta \overline{B_{tot}}} \overline{B_{tot}}.
	\label{eq: A3}
\end{equation}
Since the magnetic field is dominated by random fields, the average total field is small ($\overline{B_{tot}} \rightarrow 0$), rendering the second term of RHS in Eq. \ref{eq: A3} small.
Eq. \ref{eq: A3} indicates that the collective field at Rossby-scale ($\overline{B_{tot}}$) is not large enough to alter the structure of the random fields ($\widetilde{B} \rightarrow 0$).
Thus, we can approximate the total magnetic field as the small-scale stochastic field $\overline{B_{tot}} \sim \overline{B_{st}}$. 
This suggests that the perturbation of the Rossby wave has a minor influence on random fields.
So, the averaged magnetic stress tensor remains unchanged:
\begin{equation}
	\overline{B_{tot}^2} = \overline{\; (B_{st} + \frac{\delta B_{st}}{\delta \overline{B_{tot}}} \overline{B_{tot}})^2 \;} \simeq \overline{B_{st}^2}.
	\label{eq:bbbb}
\end{equation}
This indicates that the random field energy is fixed under the influence of the Rossby turbulence, as described by the random-field average method. 
Thus, one can simplify the calculation by ignoring the perturbation of random fields $\widetilde{B}$.

\bibliographystyle{apj}
\bibliography{main.bib}



\end{document}